\documentclass[prd,aps,showpacs,nofootinbib,floats,floatfix,superscriptaddress,
twocolumn,letterpaper]{revtex4}
\usepackage{psfrag}
\usepackage[dvips]{graphicx}
\usepackage{amsmath}
\usepackage{amsfonts}
\usepackage{color}
\usepackage{dcolumn}
\usepackage{bm}
\usepackage{amsmath}

\begin{document}

\title{Relating gravitational wave constraints from primordial 
  nucleosynthesis, pulsar timing, laser interferometers, and the CMB:
  implications for the early universe}

\author{Latham A. Boyle} 

\affiliation{Canadian Institute for Theoretical Astrophysics
  (CITA), University of Toronto, 60 St.~George Street, Toronto,
  Ontario, M5S 3H8, Canada}

\affiliation{Department of Physics, Princeton University, Princeton,
  New Jersey 08544, U.S.A.}

\author{Alessandra Buonanno}

\affiliation{Maryland Center for Fundamental Physics, 
  Department of Physics, University of Maryland, College Park,
  MD 20742, U.S.A.}

\date{August 2007}
              
\begin{abstract}
  We derive a general master equation relating the gravitational-wave
  observables $r$ and $\Omega_{0}^{{\rm gw}}(f)$; or the observables
  $\Omega_{0}^{{\rm gw}}(f_{1}^{})$ and $\Omega_{0}^{{\rm
      gw}}(f_{2}^{})$.  Here $r$ is the so-called ``tensor-to-scalar
  ratio,'' which is constrained by cosmic-microwave-background (CMB)
  experiments; and $\Omega_{0}^{{\rm gw}}(f)$ is the energy spectrum
  of primordial gravitational-waves, which is constrained {\it e.g.}\ 
  by pulsar-timing (PT) measurements, laser-interferometer (LI)
  experiments, and the standard Big Bang Nucleosynthesis (BBN) bound.
  Differentiating the master equation yields a new expression for the
  tilt $d\,{\rm ln}\,\Omega_{0}^{{\rm gw}}(f)/d\,{\rm ln}\,f$ of the
  present-day gravitational-wave spectrum.  The relationship between
  $r$ and $\Omega_{0}^{{\rm gw}}(f)$ depends sensitively on the
  uncertain physics of the early universe, and we show that this
  uncertainty may be encapsulated (in a model-independent way) by two
  quantities: $\hat{w}(f)$ and $\hat{n}_{t}^{}(f)$, where
  $\hat{n}_{t}^{}(f)$ is a certain logarithmic average over
  $n_{t}^{}(k)$ (the primordial tensor spectral index); and
  $\hat{w}(f)$ is a certain logarithmic average over $\tilde{w}(a)$
  (the effective equation-of-state parameter in the early universe,
  after horizon re-entry).  Here the {\it effective} equation-of-state
  parameter $\tilde{w}(a)$ is a combination of the {\it ordinary}
  equation-of-state parameter $w(a)$ and the bulk viscosity
  $\zeta(a)$.  Thus, by comparing observational constraints on $r$ and
  $\Omega_{0}^{{\rm gw}}(f)$, one obtains (remarkably tight)
  constraints in the $\{\hat{w}(f),\hat{n}_{t}^{}(f)\}$ plane.  In
  particular, this is the best way to constrain (or detect) the
  presence of a ``stiff'' energy component (with $w>1/3$) in the early
  universe, prior to BBN.  (The discovery of such a component would be
  no more surprising than the discovery of a tiny cosmological
  constant at late times!)  Finally, although most of our analysis
  does {\it not} assume inflation, we point out that if CMB
  experiments detect a non-zero value for $r$, then we will
  immediately obtain (as a free by-product) a new upper bound $\hat{w}
  \lesssim0.55$ on the logarithmically averaged effective
  equation-of-state parameter during the ``primordial dark age''
  between the end of inflation and the start of BBN.
\end{abstract}

\maketitle

\section{Introduction}
\label{introduction}

A variety of different experiments (some already operating, others in
various stages of development) are hoping to detect gravitational
waves (tensor perturbations) from the early universe.  In particular,
at long wavelengths, cosmic-microwave-background (CMB) experiments
\cite{QUAD, BICEP, MBI, PAPPA, CLOVER, BRAIN, POLARBEAR, QUIET,
  PLANCK, SPIDER, EBEX, CMBPOL,CMBTaskForce} will measure (or tightly
constrain) the so-called tensor-to-scalar ratio $r$ by searching for
its characteristic ``$B$-mode'' imprint in the CMB polarization
anisotropy \cite{cmb_polar, curlmodes, bmodes}.  And on shorter
wavelengths, various techniques --- including pulsar-timing (PT)
\cite{Kaspi:1994hp, Jenet:2006sv, Kramer:2004hd} and
laser-interferometer (LI) experiments \cite{LIGO, VIRGO, GEO, TAMA,
  LISA, BBO, DECIGO} --- will measure or constrain the present-day
gravitational-wave energy spectrum $\Omega_{0}^{{\rm gw}}(f)$.

The coming decade is likely to see exciting progress in this area.  At
the lowest frequencies, CMB polarization experiments will either
detect gravitational waves from inflation \cite{inflation, russians,
  AH, BA, VS, KW, turner_bound, HK, Smithetal, Boyle:2005ug}, or else
rule out the simplest (and arguably the most compelling) inflationary
models \cite{Boyle:2005ug}.  At intermediate frequencies, pulsar
timing {\it arrays} \cite{Jenet:2006sv, Kramer:2004hd} will reach far
beyond the gravitational-wave sensitivity of individual pulsars.  And
at high frequencies, the sensitivity of ground-based
gravitational-wave detectors (and also the space-based mission LISA,
if it is launched) will surpass the so-called ``standard Big Bang
Nucleosynthesis (sBBN) bound'' by several orders of magnitude, and
thus place genuinely new constraints on the primordial gravitational
wave signal at high frequencies.

Since primordial gravitational waves provide a rare and precious
window onto the extremely-high-energy physics of the infant universe,
it is essential to think carefully about the information that they
carry.

In this paper we will present a general (yet rather simple) master
equation (\ref{master_eq}) connecting the long-wavelength observable
$r$ to the short-wavelength observable $\Omega_{0}^{{\rm gw}}(f)$.
The goal is to clarify the relationship between gravitational-wave
constraints at different wavelengths, and to highlight the important
and unique information about the early universe that is encoded in
this relationship.

What exactly {\it do} we learn, in general, by comparing
long-wavelength constraints on $r$ and shorter-wavelength constraints
on $\Omega_{0}^{{\rm gw}}(f)$?  From the master equation
(\ref{master_eq}), we will see that this type of comparison should be
interpreted as primarily constraining two quantities, $\hat{w}(f)$ and
$\hat{n}_{t}^{}(f)$, which encode information about the early universe
in a model-independent way.  These two quantities are defined by
Eqs.~(\ref{def_w_hat}) -- (\ref{def_n_hat}), and explained in detail
in Sec.~\ref{master_presentation}.  For now, let us briefly discuss
their physical meaning: $\hat{n}_{t}^{}(f)$ is the logarithmic average
(over a certain range of comoving wavenumber $k$) of the primordial
tensor spectral index $n_{t}^{}(k)$; and $\hat{w}(f)$ is the
logarithmic average (over a certain range of the cosmological scale
factor $a$) of the effective equation-of-state parameter
$\tilde{w}(a)$ in the early universe ({\it after} horizon re-entry).
Here the {\it effective} equation-of-state parameter $\tilde{w}(a)$ is
a combination of the {\it ordinary} equation-of-state parameter $w(a)$
and the bulk viscosity $\zeta(a)$: see Eq.~(\ref{def_w_tilde}).  

A key advantage of our current formulation in general (and of the
variables $\hat{w}(f)$ and $\hat{n}_{t}^{}(f)$, in particular) is that
$w(a)$, $\zeta(a)$ and $n_{t}^{}(k)$ may be arbitrary functions of $a$
and $k$, respectively.  So, in particular, we will {\it not} take $w$
or $n_{t}^{}$ to be constant (or piecewise constant), as is often
assumed in analytical treatments of primordial gravitational waves.
The point is that, when deriving Eq.~(\ref{master_eq}), the quantities
$\hat{w}(f)$ and $\hat{n}_{t}^{}(f)$ naturally arise as the most
direct and general encapsulation of the uncertain early-universe
physics that enters into the relationship between $r$ and
$\Omega_{0}^{{\rm gw}}(f)$.

As an application, we will stress that comparison $r$ and
$\Omega_{0}^{{\rm gw}}(f)$ provides the most powerful way to constrain
the equation-of-state parameter $w(a)$ during the ``primordial dark
age.''  Here we use the phrase ``primordial dark age'' to refer to the
epoch separating the end of inflation from the start of Big Bang
Nucleosynthesis (BBN).  Note that, on a logarithmic scale, this
primordial dark age spans a large fraction of cosmic history: the
energy scale of BBN is $\sim 10^{-3}~{\rm GeV}$, while the energy
scale at the end of inflation may exceed $10^{16}~{\rm GeV}$.  And
yet, although there is a standard {\it theoretical} picture of how the
universe behaves during this early epoch, we currently have
essentially no direct {\it observational} constraints.

\begin{figure}
  \begin{center}
    \includegraphics[width=3.1in]{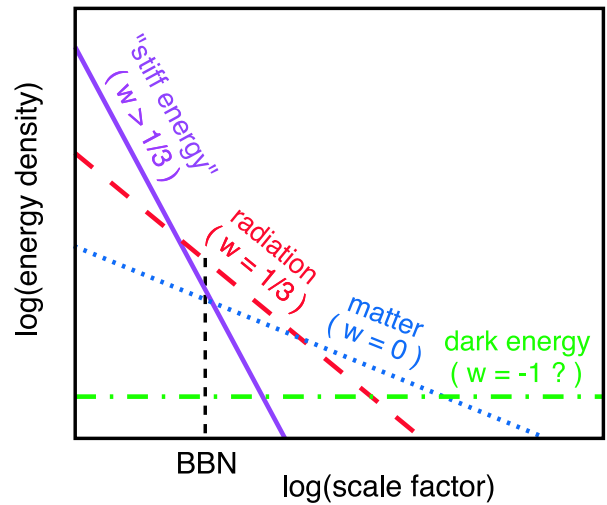}
  \end{center}
  \caption{How the components of the cosmological energy budget
    scale with cosmological expansion: ``stiff energy'' (solid purple
    line), radiation (long-dashed red line), matter (dotted blue
    line), and dark energy (dot-dashed green line).  Components with
    higher $w$ tend to dominate at earlier times.  Our universe may be
    dominated by a ``stiff energy'' component (with $w>1/3$) prior to
    Big Bang Nucleosynthesis (but after inflation).}
  \label{scaling}
\end{figure}

In fact, there are several reasons to be nervous about one of the key
(implicit) assumptions in the standard picture of the primordial dark
age: namely, the assumption that the equation-of-state satisfies
$w\leq 1/3$.  The first reason to worry is rather general: since the
energy density of a cosmological matter component scales as
$\rho\propto a^{-3(1+w)}$, components with lower $w$ dilute more
slowly.  Thus, just as an exotic component with $w$ sufficiently {\it
  low} will tend to dominate the cosmic energy budget at sufficiently
{\it late} times (think of ``dark energy'' with $w<-1/3$), an exotic
component with $w$ sufficiently {\it high} (call it ``stiff energy''
with $w>+1/3$) will tend to dominate the cosmic energy budget at
sufficiently {\it early} times (see Fig.~\ref{scaling}).
Indeed, as we look backward past BBN, the primordial dark age provides
a huge window in which a stiff energy component might overtake
radiation as the dominant component in the cosmic energy budget,
without coming into conflict with any current observational
constraints.  It is also worth noting that there are perfectly
sensible energy components with $w>1/3$ which might be present in the
early universe.  For example, a homogeneous scalar field $\phi(t)$
with vanishing (or negligible) potential energy $V(\phi)=0$ has $w=1$;
and, in fact, supergravity and string theory seem to naturally predict
many (embarrassingly many!)  scalar moduli fields with precisely this
property.  Furthermore, various authors~\cite{Peebles:1998qn,
  Giovannini:1999bh, Giovannini:1999qj, Sahni:2001qp, Tashiro:2003qp,
  Chung:2007vz} have considered inflation models in which the inflaton
field itself experiences a period of free ($w=1$) evolution at the end
of inflation; or some other equation-of-state stiffer than radiation
\cite{Grishchuk:1991kk, Gasperini:1996mf, Giovannini:1998bp,
  Giovannini:1999yy, Creighton:1999cv,Seto:2003kc}.

``Stiff ($w>1/3$) energy'' in the early universe may seem like an
exotic possibility.  But would the discovery of ``stiff energy'' at
early times be any more surprising than our apparent discovery of
``dark energy'' at late times?  One lesson that we have learned from
dark matter and dark energy is that the universe has an unmistakable
penchant for new and unexpected energy components; and it is important
to check for these components observationally, if possible, rather
than simply assuming that they are not there.  We will stress that the
comparison of constraints on $r$ and $\Omega_{0}^{{\rm gw}}(f)$
provides the best means for carrying out such a check.

One of the most important results in this paper comes from considering
the relationship between the CMB constraint on $r$ and the sBBN
constraint on $\Omega_{0}^{{\rm gw}}(f)$.  If CMB polarization
experiments succeed in detecting a non-zero value for the primordial
tensor-to-scalar ratio $r$, this will be widely interpreted as
providing evidence for inflation.  But we show that, if these
primordial tensor fluctuations are really generated by inflation, then
(in combination with the current sBBN constraint on $\Omega_{0}^{{\rm
    gw}}(f)$), this will also imply an immediate and important
supplementary result: namely a remarkably tight bound in the
$\{\hat{w}(f),\hat{n}_{t}^{}(f)\}$ plane.  This bound in the
$\{\hat{w}(f),\hat{n}_{t}^{}(f)\}$ plane is derived in
Sec.~\ref{BBN_constraints}, and shown in Fig.~\ref{wntBBNfig}.
  
If CMB polarization experiments detect a non-zero value for $r$, then
the bound depicted in Fig.~\ref{wntBBNfig} will be a qualitatively new
piece of model-independent information about the early universe ---
which is very exciting, since such information is notoriously hard to
obtain!  One way to look at the bound is as follows: If we assume that
the bulk viscosity $\zeta(a)$ is negligible after inflation, and also
that the primordial tensor power spectrum $\Delta_{h}^{2}(k)$ is
nearly flat (which is a prediction of inflation), then we obtain an
upper bound $\langle w\rangle<0.55$ on the logarithmic average of the
equation-of-state parameter $w(a)$ during the primordial dark age
separating the end of inflation from the BBN epoch.
  
It is important to clarify the range of validity of our analysis.
When we use $\Omega_{0}^{{\rm gw}}(f)$ in this paper, we are referring
only to {\it primordial} gravitational waves --- and, more
specifically, only to those gravitational waves that were generated
well {\it before} the corresponding comoving wavelength ``entered the
Hubble horizon'' ({\it i.e.}\ became shorter than the instantaneous
Hubble length).  Apart from this restriction, the results are quite
general, and make no assumptions about the physical mechanism
responsible for generating the gravitational waves.  For example, our
analysis applies to the primordial gravitational-wave spectrum
generated during inflation; and it applies equally well to the
primordial gravitational-wave spectra generated by the
``pre-big-bang''~\cite{PBB,PBBgw} and
``cyclic/ekpyrotic''~\cite{EKP,CYCLIC,EKPgw} alternatives to
inflationary cosmology; and, although all of the previous three
examples (inflationary, pre-big-bang, and ekpyrotic/cyclic cosmology)
generate primordial gravitational waves through the cosmological
amplification of quantum fluctuations, our analysis would also apply
to models that generate a primordial gravitational-waves via some
completely different mechanism (as long as they are generated {\it
  prior} to horizon entry).\footnote{A caveat is that the derivation
  of Eq.~(\ref{f_end}) specifically applies to standard inflation
  \cite{inflation} (and {\it not} to pre-big-bang \cite{PBB} or
  ekpyrotic/cyclic \cite{EKP,CYCLIC} models).  But this is a very mild
  caveat, since Eq.~(\ref{f_end}) is used only in
  Sec.~\ref{BBN_constraints}, which deals with models that produce a
  {\it detectable} value for $r$ (which pre-big-bang and
  ekpyrotic/cyclic models do not \cite{PBBgw,EKP,EKPgw}).}  On the
other hand, our analysis {\it does not} apply, {\it e.g.}\ to the
gravitational-wave spectrum produced by the incoherent superposition
of signals from merging binary stars~\cite{Farmer:2003pa}, or by a
hypothetical period of preheating after inflation~\cite{preheatingGW},
or by bubble collisions after a cosmological phase
transition~\cite{phasetransGW} --- since all of these production
mechanisms result in gravitational waves that are {\it shorter} than
the instantaneous Hubble length at the time when they are generated.

This paper is organized as follows.  Sec.~\ref{notation} introduces
some notation.  In Sec.~\ref{master_presentation} we present and
explain the master equation (\ref{master_eq}) which relates $r$ and
$\Omega_{0}^{{\rm gw}}(f)$, and will serve as the basis for most of
our analysis.  In Sec.~\ref{two_consequences} we use the master
equation (\ref{master_eq}) to derive two simple results.  The first
result is Eq.~(\ref{short_short_comparison}) which expresses the
relationship between $\Omega_{0}^{{\rm gw}}(f_{1})$ and
$\Omega_{0}^{{\rm gw}}(f_{2}^{})$ --- that is, two different
short-wavelength constraints ({\it e.g.} from LIGO and LISA) at two
different frequencies $f_{1}^{}$ and $f_{2}^{}$.  The second result is
Eq.~(\ref{energy_tilt}), which significantly generalizes previous
expressions for the tilt $d\,{\rm ln}\,\Omega_{0}^{{\rm
    gw}}(f)/d\,{\rm ln}\,k$ of the present-day energy spectrum of
primordial gravitational-waves.  In Sec.~\ref{CMB_plus_LI/PT} we
analyze the implications of combining CMB constraints on $r$ with LI
and PT constraints on $\Omega_{0}^{{\rm gw}}(f)$.  The section breaks
into 4 parts, depending on whether we suppose that CMB and LI/PT
experiments have detected the gravitational-wave background, or merely
bounded it from above.  The results are summarized in Figs.~2, 3 and
4.  In Sec.~\ref{cross_check}, we discuss the following point: If LI
or PT experiments detect an unexpectedly strong stochastic
gravitational-wave signal $\Omega_{0}^{{\rm gw}}(f)$, then there is a
rough observational consistency check that this signal should satisfy,
if it is truly of primordial origin.  In Sec.~\ref{BBN_constraints} we
analyze the constraint in the $\{\hat{w}(f), \hat{n}_{t}^{}(f)\}$
plane that follows from combining a CMB {\it detection} of $r$ with
the sBBN bound on $\Omega_{0}^{{\rm gw}}(f)$.  As mentioned above,
this constraint is rather strong; and it is also quite insensitive to
the detected value of $r$: see Fig.~5.  Finally, we conclude in
Sec.~\ref{conclusion}.  Some of the key equations in the text are
derived in appendices: in particular, Eq.~(\ref{master_eq}) is derived
in Appendix \ref{master_derivation}, and Eqs.~(\ref{f_c}) and
(\ref{f_end}) are derived in Appendix \ref{f_c_f_end_derivation}.
Appendix \ref{numbers} lists a few numbers that are useful for
converting our various algebraic expressions into numerical results
and plots.

\section{Notation}
\label{notation}

Throughout this paper, we will often use subscripts to indicate the
time at which a quantity is to be evaluated.  For example, a quantity
with subscript ``$0$'' is evaluated at the present time; a quantity
with subscript ``eq'' is evaluated at the moment of matter-radiation
equality ($\rho_{{\rm eq}}^{{\rm mat}}=\rho_{{\rm eq}}^{{\rm rad}}$);
a quantity with subscript ``$c$'' is evaluated at the redshift $z_{c}$
(defined in Sec.~\ref{master_presentation}); and a quantity with
subscript ``$k$'' is evaluated when the comoving wavenumber $k$
``re-enters the Hubble horizon'' ({\it i.e.}\ crosses from $k<aH$ to
$k>aH$).

We will also use units in which the speed of light is unity, $c=1$.

\section{The master equation}
\label{master_presentation}

Primordial gravitational wave measurements probe two basic quantities.
On long wavelengths, CMB polarization experiments constrain the
tensor-to-scalar ratio $r$.  And on shorter wavelengths, various
techniques constrain the present-day gravitational-wave energy
spectrum $\Omega_{0}^{{\rm gw}}(f)$.  In Appendix
\ref{master_derivation}, we derive a master equation relating $r$ and
$\Omega_{0}^{{\rm gw}}(f)$.  The result is:
\begin{equation}
  \label{master_eq} 
  \Omega_{0}^{{\rm gw}}(f)=\left[A_{1}A_{2}^{\hat{\alpha}(f)} 
  A_{3}^{\hat{n}_{t}^{}(f)}\right]r.
\end{equation}
As we shall see in a moment, the factor $A_{1}$ is roughly independent
of the gravitational-wave frequency $f$, while the two factors $A_{2}$
and $A_{3}$ are both proportional to $f$, so that $\Omega_{0}^{{\rm
    gw}}(f)$ is roughly proportional to
$f^{\hat{\alpha}(f)+\hat{n}_{t}^{}(f)}$.
  
Now let us carefully explain the meaning of each quantity appearing in
Eq.\ (\ref{master_eq}) --- namely, the gravitational-wave observables
$\{\Omega_{0}^{{\rm gw}}(f),r\}$, the factors $\{A_{1},A_{2},A_{3}\}$,
and the exponents $\{\hat{\alpha}(f),\hat{n}_{t}^{}(f)\}$.

The present-day gravitational-wave energy spectrum
\begin{equation}
  \Omega_{0}^{{\rm gw}}(f)\equiv\frac{1}{\rho_{0}^{{\rm crit}}}
  \frac{d\rho_{0}^{{\rm gw}}}{d\,{\rm ln}\,f}
\end{equation}
represents the present-day gravitational-wave energy density
($\rho_{0}^{{\rm gw}}$) per logarithmic frequency interval, in units
of the present-day ``critical density'' $\rho_{0}^{{\rm crit}}\equiv
3H_{0}^{2}/(8\pi G_{N}^{})$, where $H_{0}^{}$ is the present-day value
of the Hubble expansion rate, and $G_{N}^{}$ is Newton's gravitational
constant.  

The tensor-to-scalar ratio
\begin{equation}
  r\equiv \frac{\Delta_{h}^{2}(k_{{\rm cmb}})}
  {\Delta_{{\cal R}}^{2}(k_{{\rm cmb}}^{})}
\end{equation}
is the ratio of the primordial tensor power spectrum
$\Delta_{h}^{2}(k_{{\rm cmb}}^{})$ (defined in Appendix A) to the
primordial scalar power spectrum $\Delta_{{\cal R}}^{2}(k_{{\rm
    cmb}}^{})$ at the CMB wavenumber $k_{{\rm cmb}}^{}$.\footnote{
  Note that, in Eqs.~(\ref{master_eq}) and (\ref{A_1}), we could trade
  the more-commonly-used observables $r$ and $\Delta_{{\cal
      R}}^{2}(k_{{\rm cmb}}^{})$ for the single (but less commonly
  used) observable $\Delta_{h}^{2}(k_{{\rm cmb}}^{})$.}  Our
definition of the tensor-to-scalar ratio matches the convention used,
{\it e.g.}, by the WMAP experiment \cite{Peiris:2003ff,Spergel:2006hy}
and the CAMB numerical code \cite{CAMB}; but beware that there are
several alternative definitions/conventions floating around in the
literature.  The CMB wavenumber $k_{{\rm cmb}}^{}$ is the comoving
wavenumber at which CMB experiments report their constraints on
$\Delta_{{\cal R}}^{2}$, $\Delta_{h}^{2}$, and $r$: {\it e.g.}\ the
WMAP experiment \cite{Peiris:2003ff,Spergel:2006hy} uses $k_{{\rm
    cmb}}^{}/a_{0}^{} =0.002~{\rm Mpc}^{-1}$, where $a_{0}$ is the
present-day value of the cosmological scale factor.

Next consider the 3 factors $\{A_{1},A_{2},A_{3}\}$ appearing in Eq.\ 
(\ref{master_eq}).  They are given by
\begin{subequations}
  \begin{eqnarray}
  \label{A_1}
  A_{1}\!\!&\!\equiv\!&\!\!\frac{C_{2} (k)C_{3}(k)
  \Delta_{{\cal R}}^{2}(k_{{\rm cmb}}^{})\gamma}{24},\\ 
  \label{A_2}
  A_{2}\!\!&\!\equiv\!&\!\left(\frac{2\pi f}{H_{0}}\right)
  \frac{1}{(1\!+\!z_{c})\gamma^{1/2}},\\ 
  \label{A_3}
  A_{3}\!&\!\equiv\!&\!\left(\frac{2\pi f}{H_{0}}\right)
  \frac{H_{0}}{(k_{{\rm cmb}}^{}/a_{0}^{})},
  \end{eqnarray}
\end{subequations} 
where
\begin{equation}
  \label{def_Gamma}
  \gamma\equiv\frac{\Omega_{0}^{{\rm mat}}}
  {1\!+\!z_{eq}^{}}\frac{g_{\ast}^{}(\,z_{c}\,)}{g_{\ast}^{}(z_{eq})}
  \frac{g_{\ast s}^{4/3}(z_{eq})}{g_{\ast s}^{4/3}(\,z_{c}\,)}.
\end{equation}
Here $\Omega_{0}^{{\rm mat}}=\rho_{0}^{{\rm mat}}/ \rho_{0}^{{\rm
    crit}}$ is the ratio of the present-day non-relativistic matter
density $\rho_{0}^{{\rm mat}}$ to the present-day critical density
$\rho_{0}^{{\rm crit}}$.  The comoving wavenumber $k$ is related to
the physical frequency $f$ through the relation $k/a_{0}=2\pi f$.  The
cosmological scale factor $a$ is related to the cosmological redshift
$z$ through the relation $a_{0}/a=1+z$.  In particular, $z_{eq}$
denotes the redshift of matter-radiation equality ($\rho_{eq}^{{\rm
    mat}}=\rho_{eq}^{{\rm rad}}$), while $z_{c}$ denotes the highest
redshift at which we {\it know} that the universe was radiation
dominated ({\it i.e.}\ the redshift at the end of the ``primordial
dark age'' discussed in the introduction).  Given our present
observational knowledge of the early universe, it is natural to choose
$z_{c}$ to be the redshift of BBN, $z_{{\rm bbn}}^{}$; but in the
future, as our knowledge of the early universe improves, a different
choice ({\it i.e.}\ a higher redshift $z_{c}^{}$) may become more
appropriate.  The factors $g_{\ast}(z)$ and $g_{\ast s}(z)$, which
measure the effective number of relativistic degrees of freedom in the
universe at redshift $z$, are conveniently {\it defined} as follows:
If $\rho(z)$, $s(z)$, and $T(z)$ denote, respectively, the energy
density, entropy density, and temperature at redshift $z$, then
$\rho(z)=(\pi^{2}/30)g_{\ast}^{}(z)T^{4}(z)$ and $s(z)=(2\pi^{2}/45)
g_{\ast s}^{}(z)T^{3}(z)$.  For a detailed discussion of the
``correction factors'' $C_{2}(k)$ and $C_{3}(k)$, including
definitions and explicit expressions, see Appendix
\ref{master_derivation} and Ref.~\cite{Boyle:2005se}.  For now, it is
enough to note that $C_{2}(k)$ and $C_{3}(k)$ are both ${\cal O}(1)$,
which means that they will not play a very significant role in this
paper (although in other contexts they can be interesting and
important, see Ref.~\cite{Boyle:2005se}).

Finally consider the two exponents, $\hat{\alpha}(f)$ and
$\hat{n}_{t}^{}(f)$, that appear in Eq.\ (\ref{master_eq}).  The first
exponent, $\hat{\alpha}(f)$, is given by
\begin{equation}
  \hat{\alpha}(f)\equiv2\left(\frac{3\hat{w}(f)-1}{3\hat{w}(f)+1}\right),
\end{equation}
where $\hat{w}(f)$ is the logarithmic average
\begin{equation}
  \label{def_w_hat}
  \hat{w}(f)\equiv\frac{1}{{\rm ln}(a_{c}^{}/a_{k}^{})}
  \int_{a_{k}^{}}^{a_{c}^{}}\tilde{w}(a)\frac{da}{a}
\end{equation}
of the {\it effective} equation-of-state parameter $\tilde{w}(a)$ from
$a_{k}$ (the scale factor when $k=2\pi a_{0}^{}f$ re-entered the
horizon) to $a_{c}$ (the scale factor at redshift $z_{c}$).  Here the
effective equation-of-state parameter $\tilde{w}(a)$ is given by
\begin{equation}
  \label{def_w_tilde}
  \tilde{w}(a)\equiv w(a)-\frac{8\pi G_{N}\zeta(a)}{H(a)},
\end{equation}
where $w(a)=p(a)/\rho(a)$ is the {\it ordinary} equation-of-state
parameter [{\it i.e.}\ the ratio of the total cosmological pressure
$p(a)$ to the total cosmological energy density $\rho(a)=\rho_{}^{{\rm
    crit}}(a)$], $H(a)$ is the Hubble expansion rate, and $\zeta(a)$
is the bulk viscosity of the cosmological fluid (see Secs.~2.11 
and 15.11 in Ref.~\cite{WeinbergGRbook}). The second exponent,
$\hat{n}_{t}^{}(f)$, is given by the logarithmic average
\begin{equation}
  \label{def_n_hat}
  \hat{n}_{t}^{}(f)\equiv\frac{1}{{\rm ln}(k/k_{{\rm cmb}}^{})}
  \int_{k_{{\rm cmb}}^{}}^{k}n_{t}^{}(k')\frac{dk'}{k'}
\end{equation}
of the primordial tensor tilt $n_{t}^{}(k')$ over the wavenumber range
$k_{{\rm cmb}}^{}<k'<k$.  Here the primordial tensor tilt
$n_{t}^{}(k)$ is defined as the logarithmic slope of the primordial
tensor power spectrum $\Delta_{h}^{2}(k)$ at comoving wavenumber $k$:
\begin{equation}
  \label{def_nt}
  n_{t}^{}(k)\equiv\frac{d\,{\rm ln}\,\Delta_{h}^{2}(k)}{d\,{\rm ln}\,k}.
\end{equation}
We again stress that equation-of-state parameter $w(a)$ may have
arbitrary $a$-dependence, and the primordial tensor tilt $n_{t}^{}(k)$
may have arbitrary $k$-dependence.  We do {\it not} assume that $w$ or
$n_{t}^{}$ is constant.

Let us clarify the sense in which $\Omega_{0}^{{\rm gw}}(f)$ is a
``short-wavelength'' gravitational-wave observable.  We mean that, in
the master equation (\ref{master_eq}), the quantity $\Omega_{0}^{{\rm
    gw}}(f)$ represents the present-day gravitational-wave energy
spectrum {\it on scales that re-entered the Hubble horizon during the
  primordial dark age}: that is, after the end of inflation (or
whatever process produced the primordial gravitational-wave signal),
but before the redshift $z_{c}$.  In other words, the frequency $f$
that appears in equation (\ref{master_eq}) lies in the range:
\begin{equation}
  f_{c}<f<f_{end}.
\end{equation}
Here $f_{c}$ is the present-day frequency of the comoving wavenumber
$k_{c}=2\pi a_{0}^{}f_{c}$ that re-entered the Hubble horizon
($k_{c}=a_{c}H_{c}$) at redshift $z_{c}$; and $f_{{\rm end}}^{}$ is
the high-frequency cutoff of $\Omega_{0}^{{\rm gw}}(f)$.  As shown in
Appendix B, $f_{c}$ is given by
\begin{equation}
  \label{f_c}
  f_{c}=\frac{H_{0}^{}}{2\pi}(1+z_{c})\gamma^{1/2}
\end{equation}
and, if the primordial tensor spectrum is generated by inflation, then
$f_{{\rm end}}^{}$ is given by
\begin{equation}
  \label{f_end}
  f_{{\rm end}}^{}=\frac{H_{0}}{2\pi}\!\left[\!\frac{\pi^{2}r
      \Delta_{{\cal R}}^{2}(k_{{\rm cmb}}^{},\!\tau_{i}^{})}
    {16\pi G_{N}^{}H_{0}^{2}}\frac{\gamma^{1-\frac{1}{2}\hat{\alpha}}}
    {\!(1\!+\!z_{c}^{})^{\hat{\alpha}}\!}\!\left(\!\frac{a_{0}^{}H_{0}^{}}
      {k_{{\rm cmb}}^{}}\!\right)^{\!\!\hat{n}_{t}^{}}\!
  \right]^{\!1/\hat{\beta}}
\end{equation}
where, in this equation, we have used the abbreviated notation
$\{\hat{\alpha},\hat{\beta},\hat{n}_{t}^{}\}$ for the quantities
$\{\hat{\alpha}(f_{{\rm end}}^{}),\hat{\beta}(f_{{\rm end}}^{}),
\hat{n}_{t}^{}(f_{{\rm end}}^{})\}$, and defined
\begin{equation}
  \label{beta_hat}
  \hat{\beta}(f_{{\rm end}}^{})\equiv4-\hat{\alpha}(f_{{\rm end}}^{})
  -\hat{n}_{t}^{}(f_{{\rm end}}^{}).
\end{equation}
For concreteness, let us give some rough numbers: if we take
$z_{c}=z_{{\rm bbn}}^{}$ ({\it i.e.}\ the redshift at which the
temperature was $T\approx1~{\rm MeV}$), then $f_{c}=f_{{\rm
    bbn}}^{}\approx 1.8\times10^{-11}~{\rm Hz}$; and if the primordial
tensor spectrum is generated by inflation (with $\hat{n}_{t}\approx
0$), followed by a ``standard'' primordial dark age (with
$\hat{w}\approx 1/3$), then $f_{{\rm end}}^{}\approx
4.5\times10^{8}r^{1/4}~{\rm Hz}$.

Let us emphasize once again that the derivation of Eq.~(\ref{f_end})
is the {\it only} place in this paper where we assume that the
primordial gravitational wave spectrum was generated by inflation.
Since most of the results in this paper do not rely on
Eq.~(\ref{f_end}), their validity does {\it not} rely on the
correctness of inflation.  Indeed, we will only need Eq.~(\ref{f_end})
in Sec.~\ref{BBN_constraints}, when we want to combine CMB and BBN
constraints.

It is useful to interpret the master equation (\ref{master_eq}) as
follows.  From Eq.\ (\ref{master_eq}), we see that the relationship
between $r$ and $\Omega_{0}^{{\rm gw}}(f)$ is {\it much} more
sensitive to the two quantities $\{\hat{w}(f),\hat{n}_{t}^{}(f)\}$
than it is to the three quantities $\{A_{1},A_{2},A_{3}\}$ --- because
$\hat{w}(f)$ and $\hat{n}_{t}^{}(f)$ appear in the exponents of the huge
dimensionless numbers $A_{2}$ and $A_{3}$.  This means that, even
though the numerical values of $\{A_{1},A_{2},A_{3}\}$ are somewhat
uncertain (since, {\it e.g.}, $H_{0}^{}$ and $\Omega_{0}^{{\rm mat}}$
are measured with non-negligible error bars, and $C_{2}$ and $C_{3}$
are only known to be roughly equal to unity), these uncertainties do
not significantly affect the constraints on $\hat{w}(f)$ and
$\hat{n}_{t}^{}(f)$ coming from Eq.\ (\ref{master_eq}), as we shall see
in more detail below.  In other words, we may think of $\{A_{1},
A_{2}, A_{3}\}$ as ``known'' quantities; so that when we measure or
observationally constrain $r$ and $\Omega_{0}^{{\rm gw}}(f)$, the
master equation (\ref{master_eq}) allows us to directly infer
constraints on the ``unknown'' quantities $\hat{w}(f)$ and
$\hat{n}_{t}^{}(f)$.

\section{Two simple consequences}
\label{two_consequences}

Before moving on, we note two simple results that follow directly from
the master equation (\ref{master_eq}).

The first result is obtained by evaluating Eq.~(\ref{master_eq}) at
two different frequencies, $f_{1}$ and $f_{2}$, and taking the ratio
to get:
\begin{equation}
  \label{short_short_comparison}
  \frac{\Omega_{0}^{{\rm gw}}(f_{1})}{\Omega_{0}^{{\rm gw}}(f_{2})}
  =\frac{C_{2}(k_{1})C_{3}(k_{1})}{C_{2}(k_{2})C_{3}(k_{2})}
  \left[\frac{f_{1}}{f_{2}}\right]^{\hat{\alpha}(f_{1},f_{2})
  +\hat{n}_{t}^{}(f_{1},f_{2})}.
\end{equation}
Here $\hat{n}_{t}^{}(f_{1},f_{2})$ is given by
\begin{equation}
  \hat{n}_{t}^{}(f_{1},f_{2})=\frac{1}{{\rm ln}(k_{1}/k_{2})}
  \int_{k_{2}}^{k_{1}}n_{t}(k)\frac{dk}{k},
\end{equation}
where $k_{1,2}=2\pi a_{0}^{}f_{1,2}$.  And $\hat{\alpha}(f_{1},f_{2})$
is given by
\begin{equation}
  \hat{\alpha}(f_{1},f_{2})\equiv2\frac{3\hat{w}(f_{1},f_{2})-1}
  {3\hat{w}(f_{1},f_{2})+1},
\end{equation}
with
\begin{equation}
  \hat{w}(f_{1},f_{2})\equiv\frac{1}{{\rm ln}(a_{1}/a_{2})}
  \int_{a_{2}}^{a_{1}}\tilde{w}(a)\frac{da}{a},
\end{equation}
where $a_{1}^{}$ and $a_{2}^{}$ are, respectively, the values of the
scale factor when $k_{1}$ and $k_{2}$ re-entered the Hubble horizon.

Thus, whereas Eq.~(\ref{master_eq}) shows how long-wavelength (CMB)
gravitational-wave constraints relate to shorter-wavelength (pulsar,
laser-interferometer, and nucleosynthesis) constraints;
Eq.~(\ref{short_short_comparison}) explains how two shorter-wavelength
constraints ({\it e.g.}\ from LIGO and LISA) relate to one another.

The second result is obtained by differentiating Eq.\ 
(\ref{master_eq}), which yields a new expression for the logarithmic
tilt of the present-day energy spectrum:
\begin{eqnarray}
  \label{energy_tilt}
  \frac{d\,{\rm ln}\,\Omega_{0}^{{\rm gw}}(f)}{d\,{\rm ln}\,f}
  \!&\!=\!&\!n_{t}^{}(k)+2\left(\frac{3\tilde{w}(k)-1}{3\tilde{w}(k)+1}\right)
  \nonumber \\
  \!&\!\!&\!+\frac{d\,{\rm ln}\,C_{2}(k)}{d\,{\rm ln}\,k}
  +\frac{d\,{\rm ln}\,C_{3}(k)}{d\,{\rm ln}\,k},
\end{eqnarray}
where $\tilde{w}(k)$ is the value of the {\it effective}
equation-of-state parameter [see Eq.\ (\ref{def_w_tilde})] when the
comoving wavenumber $k=2\pi a_{0}^{}f$ re-enters the Hubble horizon
($k=aH$).

Note that Eq.~(\ref{energy_tilt}) generalizes earlier expressions
\cite{Giovannini:1998bp, Creighton:1999cv, Seto:2003kc} for $d\,{\rm
  ln}\,\Omega_{0}^{{\rm gw}}(f)/d\,{\rm ln}\,f$, in the sense that it
includes the corrections arising from the following 3 physical
effects, if they are present at the moment when the comoving
wavenumber $k$ is re-entering the Hubble horizon ($k=aH$) in the early
universe: (i) first, the term involving $\tilde{w}(k)$ incorporates
the correction due non-negligible bulk viscosity $\zeta$ (see Eq.
(\ref{def_w_tilde})); (ii) second, the term involving $C_{2}(k)$ is
the correction arising from time-variation of the effective
equation-of-state parameter $\tilde{w}$; and (iii) the term involving
$C_{3}(k)$ is the correction due to non-negligible tensor anisotropic
stress $\pi_{ij}$.  Again, see Appendix A and Ref.~\cite{Boyle:2005se}
for more details on the correction factors $C_{2}(k)$ and $C_{3}(k)$.

Furthermore, if the primordial gravitational-wave spectrum is produced
by the amplification of vacuum fluctuations as the mode $k$ ``exits
the Hubble horizon'' in the early universe (as in inflationary,
cyclic/ekpyrotic, and pre-big-bang cosmological models), and the
equation-of-state parameter is varying sufficiently slowly as $k$
exits the horizon, then $n_{t}^{}(k)$ is given by
\begin{equation} 
  \label{nt(w)}
  n_{t}^{}(k)=3-3\left|\frac{1-w_{{\rm exit}}^{}(k)}
    {1+3w_{{\rm exit}}^{}(k)}\right|
\end{equation}
(see Eq.~(38) in Ref.~\cite{Boyle:2004gv}), where $w_{{\rm
    exit}}^{}(k)$ is the equation-of-state parameter, evaluated at the
moment when $k$ exits the Hubble horizon.  Note that Eq.~(\ref{nt(w)})
applies equally well (i) to expanding models (like inflation, where
the modes exit the Hubble horizon while the universe is expanding with
$w<-1/3$); and (ii) to contracting models (like the pre-big-bang or
cyclic/ekpyrotic models, where the modes exit the horizon while the
universe is contracting with $w>-1/3$).

\section{CMB + LI/PT constraints}
\label{CMB_plus_LI/PT}

In this section, we explore some of the implications of the master
equation (\ref{master_eq}), focusing on the relationship between
cosmic-microwave-background (CMB) polarization experiments at long
wavelengths and laser-interferometer (LI) and pulsar-timing (PT)
experiments at shorter wavelengths.  The discussion naturally breaks
into $2\times2=4$ cases, depending on: (i) whether or not CMB
polarization experiments have successfully detected $r$, and (ii)
whether or not LI or PT experiments have successfully detected
$\Omega_{0}^{{\rm gw}}(f)$.  We number these cases as shown in Table
\ref{case_numbers}, and consider each case in turn.
\begin{table}
\begin{tabular}{|l|l|l|}
\hline
& LI/PT non-detection & LI/PT detection \\
\hline
CMB non-detection & Case 1 & Case 3 \\
\hline
CMB detection & Case 4 & Case 2 \\
\hline
\end{tabular}
\caption{The analysis in Sec.~\ref{CMB_plus_LI/PT} breaks into 4 cases, 
  depending on: {\it (i)} whether or not cosmic-microwave-background 
  (CMB) experiments have already detected a non-zero value for $r$;
  and {\it (ii)} whether or not laser-interferometer (LI) or 
  pulsar-timing (PT) experiments have already detected a non-zero 
  value for $\Omega_{0}^{{\rm gw}}(f)$.}
\label{case_numbers}
\end{table}

\subsection{Case 1: neither $r$ nor $\Omega_{0}^{{\rm gw}}(f)$ is 
  detected}
\label{Case_1}

First suppose that CMB experiments have not yet detected $r$; and that
LI/PT experiments have not yet detected $\Omega_{0}^{{\rm gw}}(f)$.
Of course, this is the current situation in 2007.

CMB observations provide an upper bound $r\leq r_{{\rm max}}^{}$.
Currently $r_{{\rm max}}^{}\approx0.5$ \cite{Spergel:2006hy}.  It is
often claimed that this long-wavelength bound implies an upper bound
on $\Omega_{0}^{{\rm gw}}(f)$ at shorter wavelengths.  Let us examine
this claim.

In fact, from the master equation (\ref{master_eq}), we see that the
upper bound is:
\begin{equation}
  \label{Omega_gw_upper_bound}
  \Omega_{0}^{{\rm gw}}(f)\leq\left[
    A_{1}A_{2}^{\hat{\alpha}_{{\rm max}}^{}}
    A_{3}^{\hat{n}_{t,{\rm max}}^{}}\right]r_{{\rm max}}^{},
\end{equation}
where 
\begin{equation}
  \label{alpha_max_from_w_max}
  \hat{\alpha}_{{\rm max}}^{}=2\left(\frac{3\hat{w}_{{\rm max}}^{}-1}
  {3\hat{w}_{{\rm max}}^{}+1}\right).
\end{equation}
In other words, in order to infer an upper bound on $\Omega_{0}^{{\rm
    gw}}(f)$ from the CMB upper bound $r\leq r_{{\rm max}}^{}$, we
need to {\it assume} two additional bounds: $\hat{w}(f)\leq
\hat{w}_{{\rm max}}^{}$ and $\hat{n}_{t}^{}(f)\leq\hat{n}_{t,{\rm
    max}}^{}$.  But these two additional bounds are theoretical
speculations about the early universe --- {\it not observational
  facts} --- so they should make us nervous.  Furthermore, since
$A_{2}$ and $A_{3}$ are huge dimensionless numbers, we see that the
upper bound on $\Omega_{0}^{{\rm gw}}(f)$ is very sensitive to the
assumed values for $\hat{w}_{{\rm max}}^{}$ and $\hat{n}_{t,{\rm
    max}}^{}$.

Now let use consider the most reasonable {\it assumptions} about
$\hat{w}_{{\rm max}}^{}$ and $\hat{n}_{t,{\rm max}}^{}$, given our
current theoretical understanding of the early universe.

What is the most reasonable assumption for $\hat{w}_{{\rm max}}^{}$?
First note that, if we assume the bulk viscosity $\zeta(a)$ is
non-negative (as required by the second law of thermodynamics), and we
assume that the equation-of-state $w(a)$ satisfies the upper bound
$w(a)\leq w_{{\rm max}}^{}$, then Eqs.\ (\ref{def_w_hat}) and
(\ref{def_w_tilde}) imply that $\hat{w}(f)$ satisfies the {\it same}
upper bound: $\hat{w}_{{\rm max}}^{}=w_{{\rm max}}^{}$.  Next note
that a fluid of massless (or extremely-relativistic) non-interacting
particles satisfies $w=1/3$; and if we give some of these particles
finite masses, or finite interactions, this tends to {\it decrease}
$w$ below $1/3$ (see Refs.~\cite{Kolb:1990vq,Boyle:2005se}).  And in
standard reheating/preheating after inflation, one also typically
finds $w\leq 1/3$ \cite{Podolsky:2005bw}.  For these reasons, and
others,
\begin{equation}
  \label{w_hat_max_best_guess}
  \hat{w}_{{\rm max}}^{}=1/3
\end{equation}
is probably the best guess.  But, as argued in
Sec.~\ref{introduction}, there are also perfectly reasonable matter
components with $w>1/3$, and there are even reasons to suspect that
these components might generically be important at sufficiently early
times (see Fig.~\ref{scaling}).  Given our current understanding of
the early universe, $\hat{w}_{{\rm max}}^{}=1/3$ is a good guess ---
but it is only a guess, and should be checked experimentally.

\begin{figure}
  \begin{center}
    \includegraphics[width=3.1in]{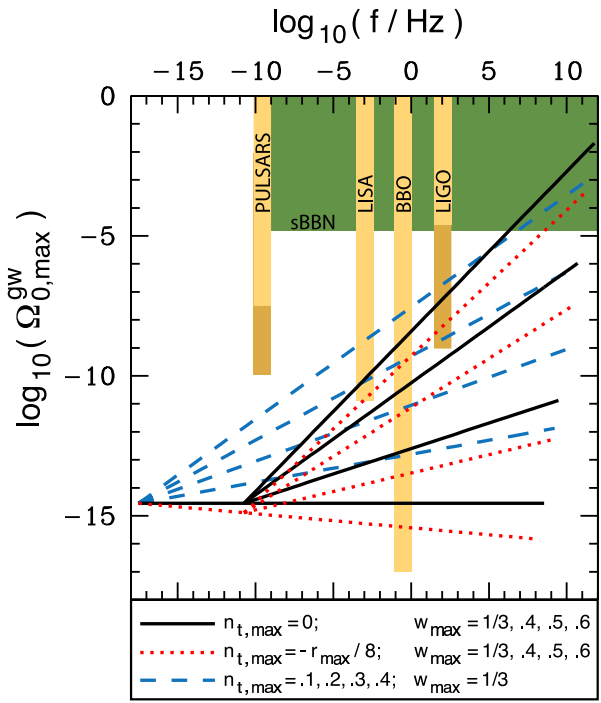}
  \end{center}
  \caption{This figure relates to ``Case 1,'' discussed in 
    Sec.~\ref{Case_1}.  The curves show the upper bound on
    $\Omega_{0}^{{\rm gw}}(f)$, over the range $f_{{\rm
        cmb}}^{}<f<f_{{\rm end}}^{}$, for various assumed values of
    $\hat{w}_{{\rm max}}^{}$ and $\hat{n}_{t,{\rm max}}^{}$.  The 4
    solid black curves correspond (from bottom to top) to
    $\hat{w}_{{\rm max}}^{}=\{1/3, 0.4, 0.5, 0.6\}$ and
    $\hat{n}_{t,{\rm max}}^{}=0$.  The 4 dotted red curves show the
    same thing, but now with $n_{t,{\rm max}}^{}=-r_{{\rm max}}^{}/8$.
    The 4 dashed blue curves correspond (from bottom to top) to
    $\hat{n}_{t,{\rm max}}^{}=\{0.1,0.2,0.3,0.4\}$ and $\hat{w}_{{\rm
        max}}^{}=1/3$.  Note that frequencies below $f_{c}^{} =f_{{\rm
        bbn}}^{}\approx 10^{-11}~{\rm Hz}$ re-entered the Hubble
    horizon {\it after} BBN, and hence are unaffected by assumptions
    about $\hat{w}$ during the primordial dark age.  The current and
    future experimental constraints shown in the figure are discussed
    in the text, at the end of Sec.~\ref{Case_1}.}
  \label{OmegaUpperBound}
\end{figure}

What is the most reasonable assumption for $\hat{n}_{t,{\rm max}}^{}$?
First note that, if we assume that the primordial tensor tilt
$n_{t}^{}(k)$ satisfies the upper bound $n_{t}^{}(k)\leq n_{t,{\rm
    max}}^{}$, then Eq.~(\ref{def_n_hat}) implies that
$\hat{n}_{t}^{}(f)$ satisfies the same upper bound: $\hat{n}_{t,{\rm
    max}}^{}=n_{t,{\rm max}}^{}$.  If we assume that the primordial
gravitational-wave spectrum was generated by inflation, then the
primordial tensor tilt is given by the well-known formula
$n_{t}^{}(k)=-2\epsilon(k)$, where $\epsilon(k)$ refers to the value
of the parameter $\epsilon(k)\equiv(3/2)(1+w_{\rm exit}(k))= -\left.
  d({\rm ln}\,H)/d({\rm ln}\,a)\right|_{k}$ when the mode $k$ leaves
the Hubble horizon ($k=aH$) during inflation.  Then, as long as the
stress-energy tensor $T_{\mu\nu}$ during inflation satisfies the
so-called ``weak energy condition'' (which, as its name suggests, is a
very mild assumption, corresponding to $w\geq-1$ in a
Friedmann-Lemaitre-Robertson-Walker (FLRW) universe), we can infer
$n_{t}^{}(k)\leq 0$.  For these reasons and others,
\begin{equation}
  \label{nt_hat_max_best_guess}
  \hat{n}_{t,{\rm max}}^{}=0
\end{equation}
is probably the best guess.  Note that this conclusion is rather
general within the context of inflation, in the sense that we have not
made reference to scalar fields, or any other details of the
(currently unknown) matter content driving inflation.  Indeed, the
conclusion should be valid as long as the following two conditions
hold: (i) gravity may be described (at least effectively) by
4-dimensional general relativity during inflation; and (ii) $\epsilon$
is $\ll 1$ and slowly varying during inflation.  Both of these
conditions are indeed satisfied by most viable inflationary models
that have been considered (single-field, multi-field, \ldots),
although there are also exotic inflationary models in the literature
that can achieve $n_{t}^{}>0$, either by violating the weak-energy
condition \cite{Baldi:2005gk} or by modifying gravity
\cite{Khoury:2006fg}.  Furthermore, although the upper bound
$\hat{n}_{t,{\rm max}}^{}=0$ applies to inflationary cosmology, it
does {\it not} apply to other cosmological models in which the
perturbations are produced during a contracting phase ({\it e.g.}\ 
``pre-big-bang'' cosmological models, which predict $n_{t}^{}=3$
\cite{PBBgw}, or cyclic/ekpyrotic models, which predict
$n_{t}^{}\approx2$ \cite{EKP,EKPgw}).
  
It is perhaps worth adding that, instead of considering inflation in
general terms, one may wish to focus on single-field inflation.  After
all, in 2007, the simplest single-field inflation models ({\it e.g.}\ 
the quadratic inflaton potential $V(\phi)= (1/2)m^{2}\phi^{2}$)
continue to agree beautifully with the current cosmological data sets
\cite{Spergel:2006hy}, and arguably provide the simplest and most
compelling available explanation of those data sets.  Since
single-field models satisfy the well known ``inflationary consistency
relation'' $n_{t}^{}(k_{{\rm cmb}}^{})=-r/8$, it turns out that we can
make the substitution $\hat{n}_{t,{\rm max}}^{}\to-r_{{\rm max}}^{}/8$
in the upper bound (\ref{Omega_gw_upper_bound}), and thereby obtain a
somewhat stronger upper bound that is still obeyed by nearly all
single-field inflationary models.

To stress that the upper bound on $\Omega_{0}^{{\rm gw}}(f)$ at high
frequencies is very sensitive to the assumed values for $\hat{w}_{{\rm
    max}}^{}$ and $\hat{n}_{t,{\rm max}}^{}$, we plot this upper bound
in Fig.~\ref{OmegaUpperBound}, for various choices of $\hat{w}_{{\rm
    max}}^{}$ and $\hat{n}_{t,{\rm max}}^{}$.
  
Fig.~\ref{OmegaUpperBound} also shows the bounds and sensitivities
from various current and future gravitational-wave constraints.  The
LIGO experiment is currently operating at its design sensitivity, and
has placed an upper bound $\Omega_{0}^{{\rm gw}}(f)<6.5\times 10^{-5}$
on the stochastic gravitational-wave background at frequencies near
$f\sim10^{2}~{\rm Hz}$ \cite{Abbott:2006zx}.  The LIGO sensitivity is
expected to increase by another factor of 10-100 within the next year
or so \cite{Abbott:2006zx}.  Then, within the next ten years, Advanced
LIGO/VIRGO is expected to reach a sensitivity of $\Omega_{0}^{{\rm
    gw}}(f)\approx 10^{-9}$ -- $10^{-8}$ \cite{Abbott:2006zx}; and
subsequent generations of ground-based LI experiments may do even
better.  LISA (the first-generation space-based LI experiment) is
expected to achieve a sensitivity of $\Omega_{0}^{{\rm
    gw}}(f)\approx10^{-11}$ at frequencies near $f\sim10^{-3}~{\rm
  Hz}$ \cite{UV}; and BBO (the second-generation space-based LI
experiment, which is specifically designed to detect a stochastic
gravitational-wave background) may be able to reach a sensitivity of
$\Omega_{0}^{{\rm gw}}(f)\approx10^{-17}$ at frequencies near
$f\sim0.3~{\rm Hz}$ \cite{BBO,Cutler:2005qq}.  Pulsar-timing
experiments have currently placed an upper bound $\Omega_{0}^{{\rm
    gw}}(f)<2\times10^{-8}$ at frequencies between $10^{-9}$ and
$10^{-8}~{\rm Hz}$ \cite{Jenet:2006sv}.  In the coming years, the
Parkes Pulsar Timing Array (PPTA), which is already operating, should
reach a sensitivity of $\Omega_{0}^{{\rm gw}}(f)\approx10^{-10}$ or
better at these frequencies \cite{Jenet:2006sv}; and in the future,
the proposed Square Kilometer Array (SKA) experiment may improve this
sensitivity by another order of magnitude or more
\cite{Kramer:2004hd}.  Finally, if short-wavelength primordial
gravitational waves had too much energy density, they would spoil the
successful predictions of BBN; so we obtain the standard Big Bang
Nucleosynthesis (sBBN) constraint \cite{Schwartzmann,
  ZeldovichNovikov, Carr1980, Allen:1996vm, Maggiore:1999vm,
  Cyburt:2004yc} depicted in Fig.~\ref{OmegaUpperBound}, and further
discussed in Sec.~\ref{BBN_constraints}.

\subsection{Case 2: both $r$ and $\Omega_{0}^{{\rm gw}}(f)$
  are detected}
\label{Case_2}

\begin{figure}
  \begin{center}
    \includegraphics[width=3.1in]{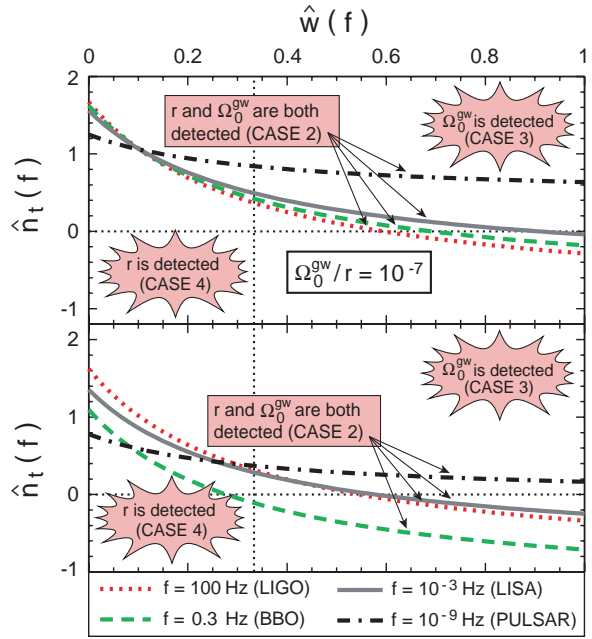}
  \end{center}
  \caption{Bounds from combining CMB and LI/PT experiments.  (This 
    figure relates to Cases 2, 3, and 4, discussed in
    Secs.~\ref{Case_2}, \ref{Case_3}, and \ref{Case_4}.)  We show
    examples of the constraints in the $\{\hat{w}(f),
    \hat{n}_{t}^{}(f)\}$ plane that follow from CMB constraints on $r$
    and LI/PT constraints on $\Omega_{0}^{{\rm gw}}(f)$.  In both the
    top and bottom panels, the 4 curves correspond to: $f_{{\rm
        LIGO}}^{}=100~{\rm Hz}$ (red dotted); $f_{{\rm
        BBO}}^{}=0.3~{\rm Hz}$ (green dashed); $f_{{\rm
        LISA}}^{}=10^{-3}~{\rm Hz}$ (grey solid); and $f_{{\rm
        pulsar}}^{}=10^{-9}~{\rm Hz}$ (black dot-dashed).  In the top
    panel, all 4 curves are plotted assuming $\Omega_{0}^{{\rm
        gw}}(f)/r=10^{-7}$.  So, for example, suppose CMB and LI/PT
    experiments find: (i) $\Omega_{0}^{{\rm gw}}(f)=10^{-8}$ and
    $r=0.1$ (Case 2); or (ii) $\Omega_{0}^{{\rm gw}}(f)=10^{-8}$ and
    $r<0.1$ (Case 3); or (iii) $\Omega_{0}^{{\rm gw}}(f)<10^{-8}$ and
    $r=0.1$ (Case 4).  Then $\{\hat{w}(f),\hat{n}_{t}^{}(f)\}$ must
    lie: (i) {\it on}, (ii) {\it above}, or (iii) {\it below} the
    respective curve.  The bottom panel is similar, but instead of all
    curves corresponding to $\Omega_{0}^{{\rm gw}}(f)/r=10^{-7}$, we
    take $\Omega_{0}^{{\rm gw}}(f)/r$ to be closer to the minimum
    possible value for each experiment: $10^{-8}$ (at $f_{{\rm
        LIGO}}^{}$); $10^{-10}$ (at $f_{{\rm LISA}}^{}$); $10^{-16}$
    (at $f_{{\rm BBO}}^{}$), and $10^{-10}$ (at $f_{{\rm
        pulsar}}^{}$).}
  \label{w_nt_plane}
\end{figure}

If CMB experiments succeed in detecting $r$, and one of the LI (or PT)
experiments (at frequency $f$) {\it also} succeeds in detecting
$\Omega_{0}^{{\rm gw}}(f)$, then the master equation (\ref{master_eq})
will yield a curve in the $\{\hat{w}(f),\hat{n}_{t}^{}(f)\}$ plane.
This is illustrated in Fig.~\ref{w_nt_plane}.  In fact, this curve
will be slightly ``fuzzy'' due to the non-vanishing error bars on $r$,
$\Omega_{0}^{{\rm gw}}(f)$, $A_{1}$, $A_{2}$, and $A_{3}$.

In particular, CMB polarization experiments are expected to be
sensitive to a tensor-to-scalar ratio as small as $r=10^{-2}$, or
smaller \cite{CMBTaskForce}; and the sensitivities of current and
future PT and LI experiments were discussed at the end of
Sec.~\ref{Case_1}.  In the top panel of Fig.~\ref{w_nt_plane}, we
imagine that $\Omega_{0}^{{\rm gw}}(f)/r=10^{-7}$ has been detected
--- {\it e.g.}\ $r=0.1$ has been detected in the CMB, and
$\Omega_{0}^{{\rm gw}}(f)=10^{-8}$ has been detected in one of the
LI/PT experiments --- and then we plot the corresponding constraint
curves, assuming that the detection of $\Omega_{0}^{{\rm gw}}(f)$
occurred at a frequency of $10^{2}~{\rm Hz}$ (LIGO), $0.3~{\rm Hz}$
(BBO), $10^{-3}~{\rm Hz}$ (LISA), or $10^{-9}~{\rm Hz}$ (PT).  And the
bottom panel of Fig.~\ref{w_nt_plane} illustrates the same thing,
assuming that the value of $\Omega_{0}^{{\rm gw}}(f)/r$ turns out to
be closer to the minimum possible value for each experiment: we use
$\Omega_{0}^{{\rm gw}}/r=10^{-9}/0.1=10^{-8}$ for LIGO (at
$10^{2}~{\rm Hz}$); and $\Omega_{0}^{{\rm
    gw}}/r=10^{-17}/0.1=10^{-16}$ for BBO (at $0.3~{\rm Hz}$); and
$\Omega_{0}^{{\rm gw}}/r=10^{-11}/0.1=10^{-10}$ for LISA (at
$10^{-3}~{\rm Hz}$); and $\Omega_{0}^{{\rm
    gw}}/r=10^{-11}/0.1=10^{-10}$ for SKA (at $10^{-9}~{\rm Hz}$).
Note that, in Fig.~\ref{w_nt_plane}, the frequency $f$ is different
for each LI/PT experiment: that is, LIGO places a constraint in the
$\{\hat{w}(f_{{\rm LIGO}}^{}), \hat{n}_{t}^{}(f_{{\rm LIGO}}^{})\}$
plane, while LISA places a constraint in the $\{\hat{w}(f_{{\rm
    LISA}}^{}), \hat{n}_{t}^{}(f_{{\rm LISA}}^{})\}$ plane, and so
forth.

\subsection{Case 3: $\Omega_{0}^{{\rm gw}}(f)$ is detected, but
  $r$ is {\it not} detected}
\label{Case_3}

In this section, let us suppose that one of the LI/PT experiments has
successfully detected $\Omega_{0}^{{\rm gw}}(f)$ at some frequency
$f$; while CMB experiments have only placed an upper bound on the
tensor-to-scalar ratio: $r\leq r_{{\rm max}}^{}$.  We will mention
three possible interpretations of this observational situation.

For the first interpretation, we rewrite the master equation
(\ref{master_eq}) as:
\begin{equation}
  \label{master_eq_v3}
  \Omega_{0}^{{\rm gw}}(f)\leq\big[A_{1}A_{2}^{\hat{\alpha}(f)}
  A_{3}^{\hat{n}_{t}^{}(f)}\big]r_{{\rm max}}^{}.
\end{equation}
As in Case 2, this equation defines a curve in the $\{\hat{w}(f),
\hat{n}_{t}^{}(f)\}$ plane.  But, whereas in Case 2 the parameters
$\hat{w}(f)$ and $\hat{n}_{t}^{}(f)$ were required to lie {\it on}
this line, in the present situation the parameters are required to lie
{\it above} the line (see Fig.~\ref{w_nt_plane}).

For the second interpretation, we rewrite the master equation
(\ref{master_eq}) as:
\begin{equation}
  \label{w_hat_lower_bound_Case_3}
  \hat{w}(f)\geq\hat{w}_{{\rm min}}^{}(f),
\end{equation}
where 
\begin{equation}
  \label{w_min_Case_3}
  \hat{w}_{{\rm min}}^{}(f)=\frac{1}{3}\left(
    \frac{2+\hat{\alpha}_{{\rm min}}^{}(f)}
    {2-\hat{\alpha}_{{\rm min}}^{}(f)}\right)
\end{equation}
and 
\begin{equation}
  \label{alpha_min_Case_3}
  \hat{\alpha}_{{\rm min}}^{}(f)=-\frac{{\rm ln}\big[
    A_{1}A_{3}^{\hat{n}_{t,{\rm max}}(f)}\,r_{{\rm max}}^{}/
    \Omega_{0}^{{\rm gw}}(f)\big]}{{\rm ln}[A_{2}]}.
\end{equation}
In other words: if we {\it assume} a theoretical upper bound for
$\hat{n}_{t}^{}(f)$, such as the standard inflationary assumption
$\hat{n}_{t,{\rm max}}^{}=0$ discussed in Sec.~\ref{Case_1}, then we
can infer that $\hat{w}(f)$ must exceed the lower bound $\hat{w}_{{\rm
    min}}^{}(f)$ given by Eqs.~(\ref{w_min_Case_3}) and
(\ref{alpha_min_Case_3}).  Furthermore, Eqs.\ (\ref{def_w_hat}) and
(\ref{def_w_tilde}) allow us to infer that the effective
equation-of-state parameter $\tilde{w}(a)$ must also satisfy the {\it
  same} lower bound
\begin{equation}
  \tilde{w}(a)\geq\hat{w}_{{\rm min}}^{}(f)
\end{equation}
for some non-empty subset of the range $a_{k}<a<a_{c}$.  And then, if
we assume $\zeta(a)\geq0$ (as required by the second law of
thermodynamics), we can also infer that the ordinary equation-of-state
parameter $w(a)$ must again satisfy the {\it same} lower bound
\begin{equation}
  w(a)\geq\hat{w}_{{\rm min}}^{}(f)
\end{equation}
for some non-empty subset of the range $a_{k}<a<a_{c}$.  
\begin{figure}
  \begin{center}
    \includegraphics[width=3.1in]{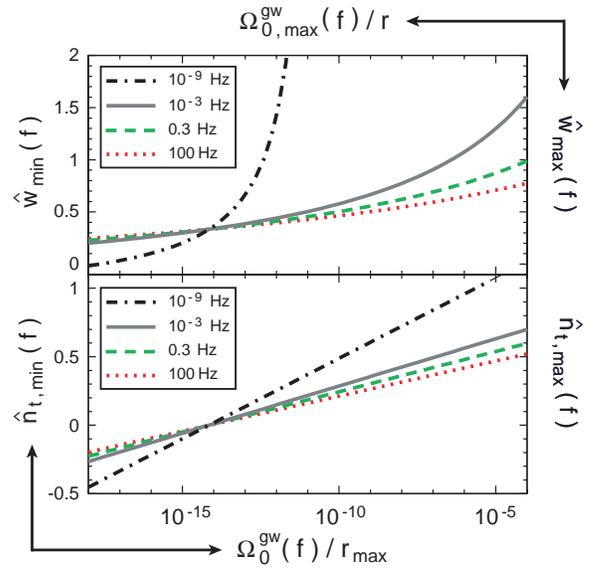}
  \end{center}
  \caption{Bounds from combining CMB and LI/PT experiments.  (This
    figure relates to Cases 3 and 4, discussed in Secs.~\ref{Case_3}
    and \ref{Case_4}.)  In both the top and bottom panels, the 4
    curves correspond to the 4 frequencies: $f_{{\rm
        LIGO}}^{}=100~{\rm Hz}$ (red dotted); $f_{{\rm
        BBO}}^{}=0.3~{\rm Hz}$ (green dashed); $f_{{\rm
        LISA}}^{}=10^{-3}~{\rm Hz}$ (grey solid); and $f_{{\rm
        pulsar}}^{}=10^{-9}~{\rm Hz}$ (black dot-dashed).  This figure
    has 2 interpretations.  In Case 3, where LI (or PT) experiments
    detect $\Omega_{0}^{{\rm gw}}(f)$ and CMB experiments obtain an
    upper bound $r_{{\rm max}}^{}$, the ``bottom'' and ``left'' axis
    labels apply, and the curves represent $\hat{w}_{{\rm min}}^{}(f)$
    (top panel, with the standard inflationary assumption
    $\hat{n}_{t,{\rm max}}^{}=0$) and $\hat{n}_{t,{\rm min}}^{}(f)$
    (bottom panel, with the standard primordial-dark-age assumption
    $\hat{w}_{{\rm max}}^{}=1/3$), so the {\it actual} values of
    $\hat{w}(f)$ and $\hat{n}_{t}^{}(f)$ lie {\it above} the curves.
    In Case 4, where CMB experiments detect $r$ and LI (or PT)
    experiments obtain an upper bound $\Omega_{0,{\rm max}}^{{\rm
        gw}}(f)$, the ``top'' and ``right'' axis labels apply, and the
    curves represent $\hat{w}_{{\rm max}}^{}(f)$ (top panel, with the
    standard inflationary assumption $\hat{n}_{t}^{}\approx0$) and
    $\hat{n}_{t,{\rm max}}^{}(f)$ (bottom panel, with the standard
    primordial-dark-age assumption $\hat{w}(f)\approx1/3$), so the
    {\it actual} values of $\hat{w}(f)$ and $\hat{n}_{t}^{}(f)$ lie
    {\it below} the curves.}
  \label{w_hat_nt_hat_vs_Omega}
\end{figure}

Eqs.~(\ref{w_min_Case_3}) and (\ref{alpha_min_Case_3}), for
$\hat{w}_{{\rm min}}^{}(f)$ as a function of $\Omega_{0}^{{\rm
    gw}}(f)/r_{{\rm max}}^{}$, are plotted in the top panel of
Fig.~\ref{w_hat_nt_hat_vs_Omega}, assuming $\hat{n}_{t,{\rm
    max}}^{}(f)=0$ (the standard inflationary assumption, discussed in
Sec.~\ref{Case_1}).  The 4 curves correspond (from top to bottom) to:
PT experiments at $f\sim10^{-9}~{\rm Hz}$ (black dot-dashed curve);
LISA at $f\sim10^{-3}~{\rm Hz}$ (grey solid curve); BBO at
$f\sim0.3~{\rm Hz}$ (green dashed curve); and LIGO at
$f\sim10^{2}~{\rm Hz}$ (red dotted curve).  Since the curves represent
$\hat{w}_{{\rm min}}^{}(f)$, the actual value of $\hat{w}(f)$ must lie
{\it above} these curves.

For the third interpretation, we rewrite the master equation
(\ref{master_eq}) as:
\begin{equation}
  \label{nt_hat_lower_bound_Case_3}
  \hat{n}_{t}^{}(f)\geq\hat{n}_{t,{\rm min}}^{}(f),
\end{equation}
where 
\begin{equation}
  \label{nt_min_Case_3}
  \hat{n}_{t,{\rm min}}^{}(f)=-\frac{{\rm ln}\big[r_{{\rm max}}^{}
    A_{1}A_{2}^{\hat{\alpha}_{{\rm max}}(f)}/\Omega_{0}^{{\rm gw}}
    (f)\big]}{{\rm ln}[A_{3}]}
\end{equation}
and
\begin{equation}
  \label{alpha_max_Case_3}
  \hat{\alpha}_{{\rm max}}^{}(f)=2\left(
    \frac{3\hat{w}_{{\rm max}}^{}(f)-1}
    {3\hat{w}_{{\rm max}}^{}(f)+1}\right).
\end{equation}
In other words: if we {\it assume} a theoretical upper bound for
$\hat{w}(f)$, such as the standard assumption $\hat{w}_{{\rm
    max}}^{}=1/3$ discussed in Sec.~\ref{Case_1}, then we can infer
that $\hat{n}_{t}^{}(f)$ must exceed the lower bound $\hat{n}_{t,{\rm
    min}}^{}(f)$ given by Eqs.~(\ref{nt_min_Case_3}) and
(\ref{alpha_max_Case_3}).  Furthermore, from Eq.\ (\ref{def_n_hat}),
we can infer that the {\it actual} primordial tensor power spectrum
$n_{t}^{}(k')$ must also satisfy the {\it same} lower bound
\begin{equation}
  n_{t}^{}(k')\geq\hat{n}_{t,{\rm min}}^{}(f)
\end{equation}
for some non-empty subset of the range $k_{{\rm cmb}}^{}<k'<k$.  

Eqs.~(\ref{nt_min_Case_3}) and (\ref{alpha_max_Case_3}), for
$\hat{n}_{t,{\rm min}}^{}(f)$ as a function of $\Omega_{0}^{{\rm
    gw}}(f)/r_{{\rm max}}^{}$, are plotted in the bottom panel of
Fig.~\ref{w_hat_nt_hat_vs_Omega}, assuming $\hat{w}_{{\rm
    max}}^{}(f)=1/3$ (a standard assumption about the primordial dark
age, as discussed in Sec.~\ref{Case_1}).  Again, the 4 curves
correspond (from top to bottom) to: PT experiments at
$f\sim10^{-9}~{\rm Hz}$ (black dot-dashed curve); LISA at
$f\sim10^{-3}~{\rm Hz}$ (grey solid curve); BBO at $f\sim0.3~{\rm Hz}$
(green dashed curve); and LIGO at $f\sim10^{2}~{\rm Hz}$ (red dotted
curve).  Since the curves represent $\hat{n}_{t,{\rm min}}^{}(f)$, the
actual value of $\hat{n}_{t}^{}(f)$ must lie {\it above} these curves.

\subsection{Case 4: $r$ is detected, but $\Omega_{0}^{{\rm gw}}(f)$
is {\it not} detected}
\label{Case_4}

Finally, in this section, let us suppose that CMB experiments have
successfully detected a non-zero value for $r$, but LI/PT experiments
have only managed to place an observational upper bound
$\Omega_{0}^{{\rm gw}}(f)<\Omega_{0,{\rm max}}^{{\rm gw}}(f)$ at
frequency $f$.  As in the previous section, we will mention three
possible interpretations of this observational situation.

For the first interpretation, we rewrite the master
equation (\ref{master_eq}) as 
\begin{equation}
  \Omega_{0,{\rm max}}^{{\rm gw}}(f)\geq\left[A_{1}
    A_{2}^{\hat{\alpha}(f)}A_{3}^{\hat{n}_{t}^{}(f)}\right]r.
\end{equation}
As in Cases 2 and 3, this equation defines a curve in the
$\{\hat{w}(f),\hat{n}_{t}^{}(f)\}$ plane.  But, whereas in Case 2 the
parameters were required to lie {\it on} this curve, and in Case 3 the
parameters were required to lie {\it above} this curve, in the present
case the parameters are required to lie {\it below} this curve (see
Fig.~\ref{w_nt_plane}).

For the second interpretation, we rewrite the master equation
(\ref{master_eq}) as:
\begin{equation}
  \label{w_hat_upper_bound_Case_4}
  \hat{w}(f)\leq\hat{w}_{{\rm max}}^{}(f),
\end{equation}
where
\begin{equation}
  \label{w_max_Case_4}
  \hat{w}_{{\rm max}}^{}(f)=\frac{1}{3}\left(
    \frac{2+\hat{\alpha}_{{\rm max}}^{}(f)}
    {2-\hat{\alpha}_{{\rm max}}^{}(f)}\right)
\end{equation}
and 
\begin{equation}
  \label{alpha_max_Case_4}
  \hat{\alpha}_{{\rm max}}^{}(f)=-\frac{{\rm ln}\big[r\,A_{1}
    A_{3}^{\hat{n}_{t}^{}(f)}/\Omega_{0,{\rm max}}^{{\rm gw}}(f)
    \big]}{{\rm ln}[A_{2}]}.
\end{equation}
In other words: if we {\it assume} assume a standard value for
$\hat{n}_{t}^{}(f)$, then we can infer that $\hat{w}(f)$ must be less
than the upper bound $\hat{w}_{{\rm max}}^{}(f)$ given by
Eqs.~(\ref{w_max_Case_4}) and (\ref{alpha_max_Case_4}).  In inflation,
the primordial gravitational wave spectrum is {\it extremely} flat, so
that the ``standard'' value may be taken as $\hat{n}_{t}^{}(f)\approx
0$.  In fact, the standard inflationary gravitational wave spectrum
has a {\it slight} negative tilt, $n_{t}^{}(k)=-2\epsilon(k)$, but it
is small enough that we can ignore it for the purpose of keeping the
present discussion simple.  It is enough to note that the slight
fuzziness in the standard inflationary value $\hat{n}_{t}^{}(f)\approx
0$ leads to slight fuzziness in the inferred upper bound
$\hat{w}_{{\rm max}}^{}(f)$.

Eqs.\ (\ref{w_max_Case_4}) and (\ref{alpha_max_Case_4}), for
$\hat{w}_{{\rm max}}^{}(f)$ as a function of $\Omega_{0,{\rm
    max}}^{{\rm gw}}(f)/r$, are plotted in the top panel of
Fig.~\ref{w_hat_nt_hat_vs_Omega}, assuming the standard inflationary
value $\hat{n}_{t}^{}(f)\approx0$.  The 4 different curves correspond
to the different LI/PT frequency bands, as already described for Case
3 in Sec.~\ref{Case_3}.  But, in Case 3, these curves represented
$\hat{w}_{{\rm min}}^{}(f)$, so that the actual value of $\hat{w}(f)$
was required to lie {\it above} the curves.  And now, in Case 4, these
same curves represent $\hat{w}_{{\rm max}}^{}(f)$, so that the actual
value of $\hat{w}(f)$ is required to lie {\it below} the curves.

For the third interpretation, we rewrite the master equation
(\ref{master_eq}) as:
\begin{equation}
  \label{nt_hat_upper_bound_Case_4}
  \hat{n}_{t}^{}(f)\leq\hat{n}_{t,{\rm max}}^{}(f),
\end{equation}
where
\begin{equation}
  \label{nt_max_Case_4}
  \hat{n}_{t,{\rm max}}^{}(f)=-\frac{{\rm ln}\big[r\,A_{1}
  A_{2}^{\hat{\alpha}(f)}/\Omega_{0,{\rm max}}^{{\rm gw}}(f)
  \big]}{{\rm ln}[A_{3}]}
\end{equation}
and 
\begin{equation}
  \label{alpha_min_Case_4}
  \hat{\alpha}(f)=2\left(\frac{3\hat{w}(f)-1}{3\hat{w}(f)+1}\right).
\end{equation}
In other words: if we {\it assume} a standard value for $\hat{w}(f)$,
then we can infer that $\hat{n}_{t}^{}(f)$ must be less than the upper
bound $\hat{n}_{t,{\rm max}}^{}(f)$ given by
Eqs.~(\ref{nt_max_Case_4}) and (\ref{alpha_min_Case_4}).  The most
common picture of the post-inflationary universe is that, after
reheating completes, the universe settles quickly into ordinary
quasi-adiabatic radiation-like expansion \cite{Podolsky:2005bw}, so
the ``standard'' value may be taken as $\hat{w}(f)\approx1/3$.  In
fact, even during standard quasi-adiabatic radiation-like expansion,
various effects --- notably conformal anomalies
\cite{Davoudiasl:2004gf, Boyle:2005se} and the evolution of
$g_{\ast}^{}$ and $g_{\ast s}^{}$ with time \cite{Kolb:1990vq} ---
cause $w$ to drop {\it slightly} below $1/3$, but these corrections
are usually small enough that we can ignore them for the purposes of
keeping the present discussion simple.  It is enough to note that the
slight fuzziness in the standard value $\hat{w}(f)\approx 1/3$ leads
to a slight fuzziness in the inferred upper bound $\hat{n}_{t,{\rm
    max}}^{}(f)$.

Eqs.\ (\ref{nt_max_Case_4}) and (\ref{alpha_min_Case_4}), for
$\hat{n}_{t,{\rm max}}^{}(f)$ as a function of $\Omega_{0,{\rm
    max}}^{{\rm gw}}(f)/r$, are plotted in the bottom panel of
Fig.~\ref{w_hat_nt_hat_vs_Omega}, assuming a ``standard'' primordial
dark age: $\hat{w}(f)\approx1/3$.  The 4 different curves correspond
to the different LI/PT frequency bands, as before.  But, in Case 3,
these curves represented $\hat{n}_{t,{\rm min}}^{}(f)$, so that the
actual value of $\hat{n}_{t}^{}(f)$ was required to lie {\it above}
the curves.  And now, in Case 4, these same curves represent
$\hat{n}_{t,{\rm max}}^{}(f)$, so that the actual value of
$\hat{n}_{t}^{}(f)$ is required to lie {\it below} the curves.

\section{Observational consistency check}
\label{cross_check}

Suppose that a pulsar-timing experiment, or a laser-interferometer
experiment like LIGO or LISA, detects a non-zero value for
$\Omega_{0}^{{\rm gw}}(f)$ that is far above the {\it expected} upper
bound $\sim10^{-15}$ which follows from assuming ``standard''
inflation plus a ``standard'' primordial dark age (see
Sec.~\ref{Case_1} and Fig.~\ref{OmegaUpperBound}).  If we wish to
interpret this as a detection of the {\it primordial}
gravitational-wave background, then we should expect it to satisfy the
following rough consistency check.

If the unexpectedly high value of $\Omega_{0}^{{\rm gw}}(f)$ is really
due to an unexpectedly high value of $\hat{w}(f)$, or an unexpectedly
high value of $\hat{n}_{t}^{}(f)$, or both, then $\Omega_{0}^{{\rm
    gw}}(f)$ should be very ``blue,'' {\it i.e.}\, rapidly rising with
frequency.  This point should be intuitively clear from a glance at
Fig.~\ref{OmegaUpperBound}, but let us be a bit more quantitative.
The {\it standard} expectation is that all four terms on the
right-hand-side of Eq.\ (\ref{energy_tilt}) are nearly zero, and hence
$\Omega_{0}^{{\rm gw}}(f)$ is nearly frequency-independent.  But if
the detected signal is actually due to an unexpectedly high value of
$\hat{w}(f)$, then the first term dominates the right-hand-side of
Eq.\ (\ref{energy_tilt}), and we expect
\begin{equation}
  \label{tilt_bound_1}
  \frac{d\,{\rm ln}\,\Omega_{0}^{{\rm gw}}}{d\,{\rm ln}\,f}
  \gtrsim 2\left(\frac{3\hat{w}_{{\rm min}}^{}-1}
    {3\hat{w}_{{\rm min}}^{}+1}\right),
\end{equation}
where $\hat{w}_{{\rm min}}^{}$ is given by Eqs.\ (\ref{w_min_Case_3})
and (\ref{alpha_min_Case_3}).  And, similarly, if the detected signal
is actually due to an unexpectedly high value of $\hat{n}_{t}^{}(f)$,
then the second term dominates the right-hand-side of Eq.\ 
(\ref{energy_tilt}), and we expect:
\begin{equation}
  \label{tilt_bound_2}
  \frac{d\,{\rm ln}\,\Omega_{0}^{{\rm gw}}}{d\,{\rm ln}\,f}
  \gtrsim \hat{n}_{t,{\rm min}}^{},
\end{equation}
where $\hat{n}_{t,{\rm min}}^{}$ is given by Eqs.\ 
(\ref{nt_min_Case_3}) and (\ref{alpha_max_Case_3}).  These
expectations can be checked within the frequency band of a single
experiment, or by comparing two different interferometers with two
separated frequency bands (like LIGO and LISA).

Note that this is just a consistency check -- it does {\it not} rule
out the possibility that the detected gravitational-wave signal is
produced by some other source, such as a cosmological phase
transition, cosmic strings, or an unanticipated astrophysical source.
Furthermore, we have been careful to use the term ``expect'' rather
than ``predict'' in this section, since it should be clear that Eqs.\ 
(\ref{tilt_bound_1}) and (\ref{tilt_bound_2}) are {\it not} firm
predictions.  Nevertheless, they are sufficiently strong expectations
that --- depending on whether or not they are confirmed --- they could
significantly increase or decrease our confidence in the ``Case 2'' or
``Case 3'' interpretations discussed in Secs.~\ref{Case_2} and
\ref{Case_3}.

\section{CMB + BBN constraints}
\label{BBN_constraints}

In this section, let us suppose that CMB experiments have succeeded in
detecting $r$, and combine this information with the well-known
``standard Big Bang Nucleosynthesis'' (sBBN) constraint 
\cite{Schwartzmann, ZeldovichNovikov, Carr1980, Allen:1996vm,
    Maggiore:1999vm, Cyburt:2004yc}: 
\begin{equation}
  \label{sBBN}
  \int_{f_{{\rm bbn}}^{}}^{f_{{\rm end}}^{}}
  \Omega_{0}^{{\rm gw}}(f)\frac{df}{f}\leq 1.5\times10^{-5}.
\end{equation}
Note that this constraint only applies to the part of the present-day
gravitational-wave spectrum that was generated {\it prior} to Big Bang
Nucleosynthesis; and the integral only runs over frequencies $f$
corresponding to comoving wavenumbers $k$ that were already ``inside
the Hubble horizon'' ($k>a_{{\rm bbn}}^{}H_{{\rm bbn}}^{}$) at the
time of Big Bang Nucleosynthesis (BBN) at photon temperature $T\sim
1~{\rm MeV}$.  In particular, the lower integration limit, $f_{{\rm
    bbn}}^{} \approx1.8\times10^{-11}~{\rm Hz}$, corresponds to the
mode that was on the Hubble horizon ($k_{{\rm bbn}}^{}=a_{{\rm
    bbn}}^{} H_{{\rm bbn}}^{}$) at the time of BBN, while the upper
integration limit $f_{{\rm end}}^{}$ corresponds to the high-frequency
cutoff of the primordial gravitational-wave spectrum.  For example, if
the primordial gravitational-wave spectrum was generated by inflation,
then the spectrum cuts off exponentially fast for $k>k_{{\rm end}}$,
where $k_{{\rm end}}^{}=a_{{\rm end}}^{}H_{{\rm end}}^{}$ is the
comoving wavenumber that was on the Hubble horizon at the end of
inflation. This corresponds to the present-day frequency $f_{{\rm
    end}}^{}$ given by Eq.~(\ref{f_end}).

Although, the sBBN constraint (\ref{sBBN}) is technically an integral
constraint (non-local in frequency space), in practice it effectively
acts as an algebraic constraint (local in frequency space) of the form
$\Omega_{0}^{{\rm gw}}(f)<1.5\times10^{-5}$ for $f_{{\rm bbn}}^{}<
f<f_{{\rm end}}$.  $\Omega_{0}^{{\rm gw}}(f)$ can only exceed this
bound by having a very narrow spike with $(\delta f)/f_{0}\ll 1$,
where $f_{0}$ is the peak of the spike, and $\delta f$ is its
characteristic width; but (as far as we are aware) there are no known
mechanisms for producing such a narrow spike in the primordial
gravitational-wave spectrum, and we will neglect this possibility.

\begin{figure}
  \begin{center}
    \includegraphics[width=3.1in]{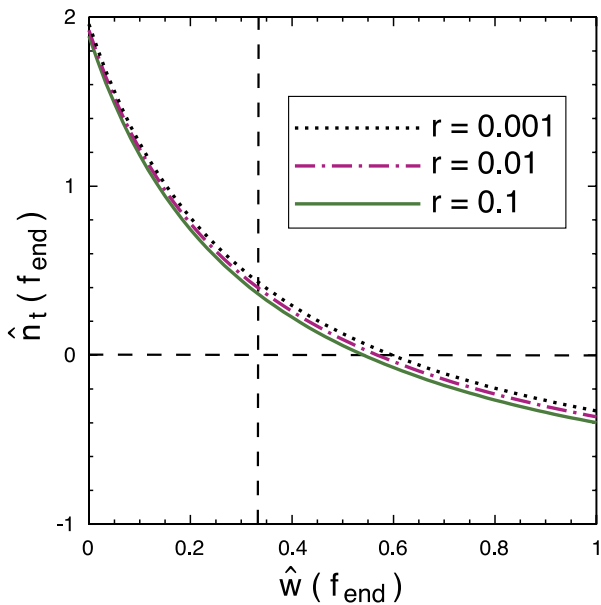}
  \end{center}
  \caption{Bound from combining sBBN and CMB constraints.  If CMB 
    experiments detect $r$, then the sBBN gravitational-wave
    constraint immediately requires $\{\hat{w}(f_{{\rm end}}^{}),
    \hat{n}_{t}^{}(f_{{\rm end}}^{})\}$ to lie {\it below} the curves
    shown in the figure.  From top to bottom, the curves correspond
    to: $r=10^{-3}$ (black dotted curve), $r=10^{-2}$ (purple
    dot-dashed curve), and $r=10^{-1}$ (green solid curve).  Note that
    the curves are very insensitive to $r$: they hardly move as $r$
    varies over the range in which it can be realistically detected by
    CMB polarization experiments ($10^{-3}<r<10^{-1}$).  The
    horizontal and vertical dashed lines point out that, for
    $r=\{10^{-1},10^{-2},10^{-3}\}$, respectively: (a) if
    $\hat{n}_{t}^{}(f)$ is assumed to take its ``standard''' value
    ($\approx 0$), then $\hat{w}(f_{{\rm end}}^{})\lesssim
    \{0.54,0.57,0.60\}$; and (b) if $\hat{w}(f)$ is assumed to take
    its ``standard'' value ($\approx 1/3$), then $\hat{n}_{t}^{}
    (f_{{\rm end}})\lesssim\{0.36,0.40,0.43\}$.}
  \label{wntBBNfig}
\end{figure}

Thus, for any frequency $f$ in the range $f_{{\rm bbn}}^{}<f<f_{{\rm
    end}}^{}$, we can directly use all of the equations from ``Case
4'' in the previous section, as long as we set $\Omega_{0,{\rm
    max}}^{{\rm gw}}(f)=1.5\times10^{-5}$ in those equations.
Furthermore, to maximize the length of the ``lever arm'' between the
CMB and BBN constraints, let us consider the case $k\to k_{{\rm
    end}}^{}$ and $f\to f_{{\rm end}}^{}$.  Then
Eqs.~(\ref{nt_max_Case_4}) and (\ref{alpha_min_Case_4}) define a curve
in the $\{\hat{w}(f_{{\rm end}}^{}), \hat{n}_{t}^{}(f_{{\rm
    end}}^{})\}$ plane (shown in Fig.~\ref{wntBBNfig}), and the actual
values of $\hat{w}(f_{{\rm end}}^{})$ and $\hat{n}_{t}^{} (f_{{\rm
    end}}^{})$ must lie {\it below} this curve.

Note, in particular, that the constraint curve hardly varies as $r$
varies over the range of realistic future detectability
$10^{-3}<r<10^{-1}$.  Furthermore, for
$r=\{10^{-1},10^{-2},10^{-3}\}$, respectively: (a) if
$\hat{n}_{t}^{}(f)$ is assumed to take its ``standard'' inflationary
value ($\hat{n}_{t}^{}(f)\approx0$, see Secs.~\ref{Case_1} and
\ref{Case_3}), then we obtain the upper bound $\hat{w}(f_{{\rm
    end}}^{})\lesssim\{0.54,0.57,0.60\}$; and (b) if $\hat{w}(f)$ is
assumed to take its ``standard'' post-inflationary value
($\hat{w}(f)\approx 1/3$, again see Secs.~\ref{Case_1} and
\ref{Case_3}), then we obtain the upper bound $\hat{n}_{t}^{}(f_{{\rm
    end}}^{})\lesssim\{0.36,0.40,0.43\}$.

These results (particularly Fig.~\ref{wntBBNfig}) are new, and
model-independent, constraints on the early universe that will take
effect as soon as CMB polarization experiments detect a non-zero value
for $r$.

\section{Conclusion}
\label{conclusion}

As far as the early universe is concerned, most people think about
upcoming cosmic-microwave-background (CMB) ``B-mode'' polarization
experiments with the following goal in mind: to measure one crucial
number, $r$, which physically corresponds to measuring the energy
density of the universe, roughly $60$ e-folds before the end of
inflation.  But these B-mode experiments will actually achieve
significantly more than this: they should also be viewed as half of a
two-pronged experiment to detect or constrain the early-universe
parameters $\hat{w}(f)$ and $\hat{n}_{t}^{}(f)$, as we have described
in detail.  (The other ``prong'' of this two-pronged experiment is a
higher-frequency gravitational-wave constraint coming from
laser-interferometer experiments, pulsar-timing measurements, or
standard BBN.)  For example, if and when CMB experiments detect a
non-zero value for $r$, they will immediately obtain a {\it
  supplementary} (and remarkably strong) constraint in the
$\{\hat{w}(f_{{\rm end}}^{}), \hat{n}_{t}^{}(f_{{\rm end}}^{})\}$
plane, as shown in Fig.~\ref{wntBBNfig}.  Since quantitative and
model-independent constraints on the early universe are notoriously
hard to obtain, and we only have a handful, the possibility of
obtaining this ``supplementary'' constraint is exciting.

We have argued that combining large-wavelength constraints on $r$
(from CMB experiments) with small-wavelength bounds on
$\Omega_{0}^{{\rm gw}}(f)$ (from LI and PT experiments, and sBBN
constraints) provides the strongest way to constrain (or detect) the
existence and properties of a possible exotic ``stiff energy''
component (with $w>1/3$) \cite{Grishchuk:1991kk, Gasperini:1996mf,
  Giovannini:1998bp, Giovannini:1999yy, Peebles:1998qn,
  Giovannini:1999bh, Giovannini:1999qj, Sahni:2001qp,
  Tashiro:2003qp,Chung:2007vz} that could have dominated the universe
for some period during the primordial dark age between the end of
inflation and the BBN epoch (see Fig.~\ref{scaling}).  

We have derived several useful and general formulae for relating
primordial gravitational-wave constraints at different frequencies,
and have shown how these relationships connect to the uncertain
physics of the early universe.  In Figs.~\ref{OmegaUpperBound},
\ref{w_nt_plane}, \ref{w_hat_nt_hat_vs_Omega} and \ref{wntBBNfig}, we
have shown the constraints that will be placed on the parameters
$\hat{w}(f)$ and $\hat{n}_{t}^{}(f)$ by combining various pairs of
gravitational-wave constraints, depending on the observational
situation (that is, depending on whether CMB and/or LI/PT experiments
detect the primordial gravitational-wave background, or only place
upper limits).

\acknowledgments

L.B.~thanks Paul Steinhardt for many insightful conversations.
A.B.~acknowledges support from NSF grant PHY-0603762 and from the
Alfred P. Sloan Foundation.

\appendix

\section{Deriving the master equation}
\label{master_derivation}

  The goal of this Appendix is to derive Eq.\ 
  (\ref{master_eq_v2}), which is equivalent to Eq.~(\ref{master_eq}),
  the master equation upon which most of the analysis in this paper
  relies.  As a useful intermediate result, we also obtain
  Eq.~(\ref{T_h}), an expression for the tensor transfer function
  $T_{h}^{}(k)$.

   The derivation is broken into 4 parts.  In the first part, we
  review some background material about cosmological gravitational
  waves, leading to the presentation of Eq.~(\ref{master_eq_v1}).  The
  second and third parts are devoted to rewriting the factors
  $T_{h}^{}(k)$ and $\Delta_{h}^{2}(k,\tau_{i})$, which appear in
  Eq.~(\ref{master_eq_v1}).  Finally, in the fourth part, we collect
  and summarize our results in Eqs.~(\ref{T_h}) and
  (\ref{master_eq_v2}).

\subsection{Background material}
 
Let us start by introducing some notation, and reviewing some basic
facts about cosmological gravitational waves (tensor perturbations
\cite{Bardeen:1980kt}).  For more details, see Sec. 2 in
Ref.~\cite{Boyle:2005se}.

Tensor metric perturbations in a spatially-flat FLRW universe are
described by the line element:
\begin{equation}
  ds^{2}=a^{2}(\tau)\left[-d\tau^{2}+(\delta_{ij}+h_{ij}({\bf x},
    \tau))d{\bf x}^{i}d{\bf x}^{j}\right],
\end{equation}
where ${\bf x}$ is a comoving spatial coordinate, $\tau$ is a
conformal time coordinate, $a(\tau)$ is the FLRW scale factor, 
and the metric perturbation $h_{ij}({\bf x},\tau)$ is transverse
($h_{ij,j}=0$) and traceless ($h_{ii}=0$).  In this Appendix we follow
the convention that repeated indices ($i$ or $j$) are summed from $1$
to $3$.

The tensor power spectrum $\Delta_{h}^{2}(k,\tau)$, which represents
the contribution by modes of comoving wavenumber $k$ to the
expectation value $\langle h_{ij}({\bf x},\tau) h_{ij}({\bf
  x},\tau)\rangle$, is defined through the equation
\begin{equation}
  \label{def_Delta_t}
  \langle h_{ij}({\bf x},\tau)h_{ij}({\bf x},\tau)\rangle
  =\int \Delta_{h}^{2}(k,\tau)\frac{dk}{k}.
\end{equation}
Note that the expectation value of the left-hand-side is
  actually independent of ${\bf x}$, since a perturbed FLRW universe
  is {\it statistically} homogeneous.

CMB and LI experiments measure $\Delta_{h}^{2}(k,\tau)$ at very
different comoving wavenumbers $k$ and very different conformal times
$\tau$.  In particular, whereas LI experiments measure the {\it
  present-day} tensor power spectrum $\Delta_{h}^{2}(k,\tau_{0})$, CMB
experiments may be thought of as measuring the {\it primordial} tensor
power spectrum $\Delta_{h}^{2}(k,\tau_{i})$.  (Here $\tau_{0}$ denotes
the present time, and $\tau_{i}$ denotes a very early time, before any
modes $k$ of interest have had a chance to re-enter the Hubble
horizon.)  And whereas LI experiments are sensitive to {\it high}
comoving wavenumbers (corresponding to length scales smaller than the
Solar System), CMB experiments are sensitive to {\it low} comoving
wavenumbers (corresponding to large length scales, comparable to the
present-day Hubble radius).  CMB constraints on the primordial scalar
and tensor power spectra, $\Delta_{{\cal R}}^{2}(k,\tau_{i})$ and
$\Delta_{h}^{2}(k,\tau_{i})$, are usually quoted at a fiducial
``pivot'' wavenumber $k_{{\rm cmb}}$ in the CMB waveband. For
  example, the WMAP experiment uses $k_{{\rm cmb}}^{}/a_{0}=0.002~{\rm
    Mpc}^{-1}$, where $a_{0}$ is the present-day ($\tau=\tau_{0}$)
  value of the FLRW scale factor.

Although it is often convenient, from a theoretical perspective, to
work with the tensor power spectrum $\Delta_{h}^{2}(k,\tau)$, LI
experiments usually report their results in terms of the present-day
($\tau=\tau_{0}$) gravitational-wave energy spectrum:
\begin{equation}
  \label{def_Omega_gw}
  \Omega_{0}^{{\rm gw}}(f)\equiv\frac{1}{\rho_{0}^{{\rm crit}}}
  \frac{d\rho_{0}^{{\rm gw}}}{d\,{\rm ln}\,f},
\end{equation}
where
\begin{equation}
  \label{f_from_k}
  f=\frac{1}{2\pi}\frac{k}{a_{0}}
\end{equation}
is the present-day physical frequency of a gravitational wave
corresponding to the comoving wavenumber $k$.  Note that the
present-day energy spectrum $\Omega_{0}^{{\rm gw}}(f)$ is related to
the present-day power spectrum $\Delta_{h}^{2}(k,\tau_{0})$ through
the equation
\begin{equation}
  \label{Omega_from_Delta}
  \Omega_{0}^{{\rm gw}}(f)=\frac{1}{12}\left[\frac{2\pi f}
    {H_{0}^{}}\right]^{2}\Delta_{h}^{2}(k,\tau_{0})
\end{equation}
(see Sec.~2 in Ref.~\cite{Boyle:2005se} for a detailed derivation).

The present-day tensor power spectrum $\Delta_{h}^{2}(k,\tau_{0})$ is
related to the primordial tensor power spectrum $\Delta_{h}^{2}
(k,\tau_{i})$ via the relation
\begin{equation}
  \label{def_T_h}
  \Delta_{h}^{2}(k,\tau_{0})=T_{h}^{}(k)\Delta_{h}^{2}(k,\tau_{i}),
\end{equation}
where this equation defines the ``tensor transfer function''
$T_{h}^{}(k)$.  Combining Eqs.\ (\ref{Omega_from_Delta}) and
(\ref{def_T_h}) we obtain
\begin{equation}
  \label{master_eq_v1} 
  \Omega_{0}^{{\rm gw}}(f)=\frac{1}{12}\left[\frac{2\pi f}
  {H_{0}^{}}\right]^{2}T_{h}^{}(k)\Delta_{h}^{2}(k,\tau_{i}).
\end{equation}
This is the master equation describing the present-day
gravitational-wave energy spectrum $\Omega_{0}^{{\rm gw}}(f)$ on LI
scales.  The rest of this section is devoted to rewriting this
equation in a more concrete and useful form.  In the next two 
sections, we re-express the two factors $T_{h}^{}(k)$ and
$\Delta_{h}^{2}(k,\tau_{i})$, respectively.

\subsection{Rewriting the factor $T_{h}^{}(k)$}
\label{transfer}

First, let us focus on rewriting the factor $T_{h}^{}(k)$.  In this
paper, we make use of the general expression for the tensor transfer
function $T_{h}^{}(k)$ derived in Ref.~\cite{Boyle:2005se}.  As
explained in Ref.~\cite{Boyle:2005se}, the tensor transfer function
$T_{h}^{}(k)$ may be factored into the form
\begin{equation}
  \label{T=C1C2C3} 
  T_{h}^{}(k)=\frac{1}{2}C_{1}(k)C_{2}(k)C_{3}(k).
\end{equation}
The overall factor of $1/2$ comes from averaging over the
  oscillatory factor ${\rm cos}^{2}(k\tau+\phi(k))$ which appears in
  the tensor transfer function but is unresolvable in any forseeable
  LI experiment \cite{Allen:1999xw}. Each of the remaining 3 factors
$\{C_{1}(k), C_{2}(k), C_{3}(k)\}$ has a simple physical meaning and
is derived in detail in Ref.~\cite{Boyle:2005se}.  Here we just quote 
a few key results.

  As we shall see, the expression (\ref{T=C1C2C3}) for
  $T_{h}^{}(k)$ is dominated by the factor $C_{1}(k)\ll 1$, while the
  other two factors, $C_{2}(k)$ and $C_{3}(k)$, represent modest
  ${\cal O}(1)$ corrections.

The factor $C_{1}(k)$ is given by
\begin{equation}
  \label{C1_v1}
  C_{1}(k)=\frac{1}{(1+z_{k}^{})^{2}}
\end{equation}
where $z_{k}^{}$ is the redshift at which the mode $k$ re-enters the
Hubble horizon ($k=aH$) after inflation.  We shall return to
  this factor below.

\begin{figure}
  \begin{center}
    \includegraphics[width=3.1in]{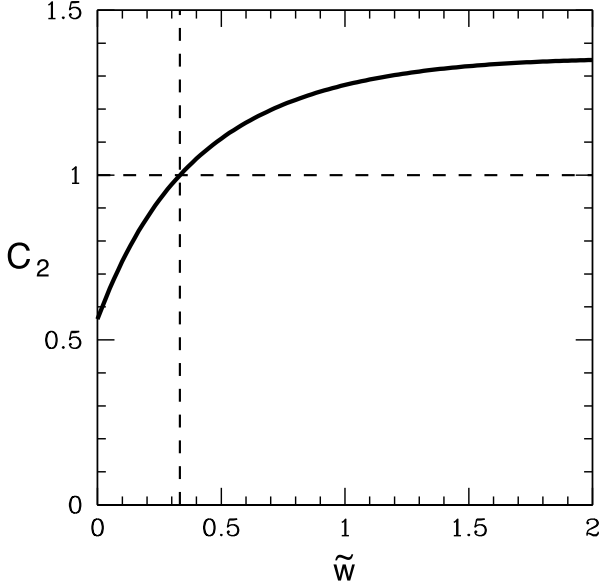}
  \end{center}
  \caption{The correction factor $C_{2}$ as a function of the 
    effective equation-of-state parameter $\tilde{w}$, as given by
    Eqs.~(\ref{C2}) and (\ref{def_alpha_k}).  Note that $C_{2}=1$ when
    $\tilde{w}=1/3$.}
  \label{C2fig}
\end{figure}

The factor $C_{2}(k)$ is given by:
\begin{equation}
  \label{C2}
  C_{2}(k)=\frac{\Gamma^{2}(\alpha_{k}^{}+1/2)}{\pi}
  \left [\frac{2}{\alpha_{k}^{}} \right ]^{2\alpha_{k}^{}}
\end{equation}
where $\Gamma(x)$ is the gamma function, and we have defined
\begin{equation}
  \label{def_alpha_k}
  \alpha_{k}^{}\equiv\frac{2}{1+3\tilde{w}_{k}^{}}.
\end{equation}
Here $\tilde{w}_{k}^{}$ is the {\it effective} equation-of-state
parameter at redshift $z_{k}^{}$, and is given by
\begin{equation}
  \label{def_w_tilde_k}
  \tilde{w}_{k}^{}=w_{k}^{}-\frac{8\pi G_{N}^{}\zeta_{k}^{}}
  {H_{k}^{}},
\end{equation}
where $w_{k}^{}\equiv p_{k}^{}/\rho_{k}^{}$ is the {\it usual}
equation-of-state parameter ({\it i.e.}\ the ratio of the total
cosmological pressure $p_{k}^{}$ to the total cosmological energy
density $\rho_{k}^{}$), $H_{k}^{}$ is the Hubble expansion rate,
$\zeta_{k}^{}$ is the bulk viscosity of the cosmological fluid (see
Secs. 2.11 and 15.11 in Ref.~\cite{WeinbergGRbook}) --- and, as
their subscripts indicate, all of these quantities are evaluated at
redshift $z_{k}^{}$.  $C_{2}$ is plotted in Fig.~\ref{C2fig}.
Note that the expression (\ref{C2}) for $C_{2}(k)$ is valid as long as
the effective equation-of-state parameter $\tilde{w}$ is not changing
rapidly relative to the instantaneous Hubble time at redshift
$z_{k}^{}$; see Ref.~\cite{Boyle:2005se} for the meaning of $C_{2}(k)$
more generally.

\begin{figure}
  \begin{center}
    \includegraphics[width=3.1in]{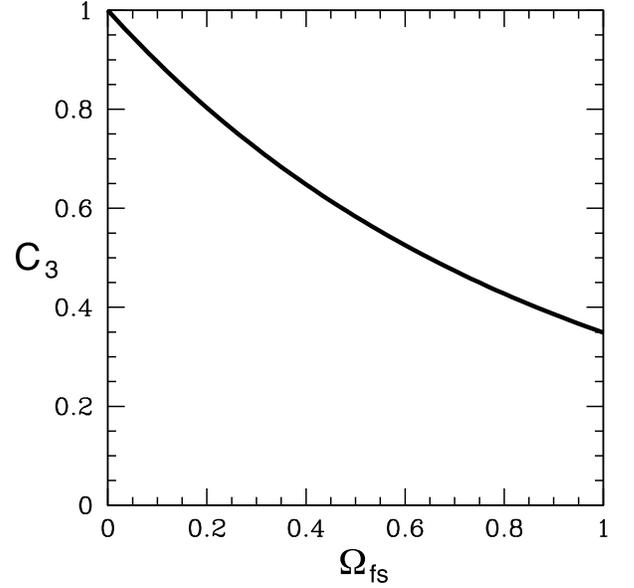}
  \end{center}
  \caption{The correction factor $C_{3}$ as a function of the 
    free-streaming fraction $\Omega_{fs}^{}=\rho_{k}^{fs}/
    \rho_{k}^{{\rm crit}}$, as given by Eqs.~(\ref{C3}) and
    (\ref{A}).}
  \label{C3fig}
\end{figure}

The factor $C_{3}(k)$ captures the modification of the primordial
gravitational-wave signal due to tensor anisotropic stress $\pi_{ij}$
({\it e.g.}\ from free-streaming relativistic particles) in the early
universe.  In particular, in the important case that the effective
equation-of-state near $z_{k}$ is radiation-like ($\tilde{w}\approx
1/3$), free-streaming relativistic particles with energy density
$\rho_{}^{fs}$ damp the gravitational wave spectrum by the factor:
\begin{equation}
  \label{C3}
  C_{3}(k)=A^{2}(k)
\end{equation}
where
\begin{equation}
  \label{A}
  A(k)\!\equiv\!-\frac{10}{7}\frac{(98\Omega_{fs}^{3}\!-\!589
    \Omega_{fs}^{2}\!+\!9380\Omega_{fs}^{}\!-\!55125)}
  {(15\!+\!4\Omega_{fs}^{})(50\!+\!4\Omega_{fs}^{})(105\!+\!4\Omega_{fs}^{})}
\end{equation}
and $\Omega_{fs}^{}\equiv\rho_{k}^{fs}/\rho_{k}^{{\rm crit}}$ is the
fraction of the critical density that is relativistically
free-streaming at redshift $z_{k}^{}$.  $C_{3}$ is plotted in
Fig.~\ref{C3fig}.

In the remainder of this section, we focus on obtaining a more
explicit expression for $(1+z_{k}^{})$, and hence for $C_{1}(k)$.  To
do this, let us proceed carefully as follows:
\begin{subequations}
  \begin{eqnarray}
    \label{F_v1a}
    \frac{k^{2}}{a_{0}^{2}H_{0}^{2}}
    \!&\!=\!&\!\frac{a_{k}^{2}H_{k}^{2}}{a_{0}^{2}H_{0}^{2}} \\
    \label{F_v1b}
    \!&\!=\!&\!\frac{1}{(1+z_{k}^{})^{2}}
    \frac{\rho_{k}^{{\rm crit}}}{\rho_{0}^{{\rm crit}}} \\
    \label{F_v1c}
    \!&\!=\!&\!\frac{1}{(1+z_{k}^{})^{2}}
    \frac{\rho_{eq}^{{\rm mat}}}{\rho_{0}^{{\rm crit}}}
    \frac{\rho_{c}^{{\rm rad}}}{\rho_{eq}^{{\rm rad}}}
    \frac{\rho_{k}^{{\rm crit}}}{\rho_{c}^{{\rm crit}}},
  \end{eqnarray}
\end{subequations}
where $\rho_{}^{{\rm mat}}$, $\rho_{}^{{\rm rad}}$, and $\rho_{}^{{\rm
    crit}}$ denote the matter density, radiation density, and critical
density, respectively.  In the first line (\ref{F_v1a}), we have used
the fact that $k=a_{k}^{}H_{k}^{}$ by definition.  In the second line
(\ref{F_v1b}), we have used the definition of the redshift $z$ to
write $a_{0}^{}/a_{k}^{}=(1+z_{k}^{})$, and the definition of the
critical density $\rho_{}^{{\rm crit}}$ to write
$H_{k}^{2}/H_{0}^{2}=\rho_{k}^{{\rm crit}}/\rho_{0}^{{\rm crit}}$.  In
the third line (\ref{F_v1c}), we have used the fact that
$\rho_{eq}^{{\rm mat}}=\rho_{eq}^{{\rm rad}}$ at the time $\tau_{eq}$
of matter-radiation equality; plus the fact that the universe is
radiation dominated at $\tau_{c}$ so that $\rho_{c}^{{\rm crit}}
=\rho_{c}^{{\rm rad}}$.

We have introduced the time $\tau_{c}$ to parametrize our threshold of
ignorance: it represents the earliest time at which we {\it know} that
the universe was already radiation dominated. But, for all we know,
the universe {\it prior} to $\tau_{c}$ may {\it not} have been
radiation-dominated: {\it e.g.}\ an exotic ``stiff'' component with
$w>1/3$ may have dominated the cosmic energy budget.  The present
agreement between the theoretical and observational understanding of
Big Bang Nucleosynthesis (BBN) strongly suggests that the universe was
already radiation dominated during BBN ({\it i.e.}\ at the time
$\tau_{{\rm bbn}}^{}$ when the temperature was $T=1~{\rm MeV}$), so it
currently makes sense to choose $\tau_{c}= \tau_{{\rm bbn}}^{}$.  But,
for the sake of generality, we leave $\tau_{c}$ unfixed in this
section, since we can easily imagine future developments --- such as
an improved understanding of primordial baryogenesis --- that would
make an earlier time $\tau_{c}\ll\tau_{{\rm bbn}}^{}$ a more
appropriate choice.  Note that for the wavenumbers of interest in this
paper --- {\it e.g.}\ those measured by laser interferometer
experiments --- the temporal ordering is
$\tau_{k}<\tau_{c}<\tau_{eq}<\tau_{0}$.

Three density ratios appear on the right-hand side of Eq.\ 
(\ref{F_v1c}).  Let us rewrite each of these in turn.  The first
density ratio is trivially rewritten as:
\begin{equation}
  \label{factor1}
  \frac{\rho_{eq}^{{\rm mat}}}{\rho_{0}^{{\rm crit}}}\!=\!
  (1+z_{eq})^{3}\Omega_{0}^{{\rm mat}},
\end{equation}  
where $\Omega_{0}^{{\rm mat}}\equiv\rho_{0}^{{\rm mat}}/
\rho_{0}^{{\rm crit}}$.

In order to rewrite the second density ratio appearing in
(\ref{F_v1c}), let us pause for a moment to review a few
  properties of an expanding bath of radiation.  The radiation bath
has energy density $\rho$ and entropy density $s$ given by (see Secs.
3.3 and 3.4 in Ref.~\cite{Kolb:1990vq}, and especially Eqs.~(3.61), (3.72)):
\begin{subequations}
  \begin{eqnarray}
    \label{rho(T)}
    \rho(z)&=&\frac{1}{30}\pi^{2}g_{\ast}^{}(z)T^{4}(z), \\
    \label{s(T)}
    s(z)&=&\frac{2}{45}\pi^{2}g_{\ast s}^{}(z)T^{3}(z),
  \end{eqnarray}
\end{subequations}
where $T(z)$ is the temperature at redshift $z$.  These
equations may be taken as the {\it definition} of the quantities
$g_{\ast}^{}(z)$ and $g_{\ast s}^{}(z)$, which represent the {\it
  effective} number of relativistic degrees of freedom in the
radiation bath at redshift $z$, as measured by the energy density
$\rho(z)$ or the entropy density $s(z)$, respectively. If the
  radiation expands quasi-adiabatically --- as is usually the case in
  the early universe --- then the entropy $a^{3}s$ remains constant
  (to a very good approximation). When this is true, Eqs.\ 
(\ref{rho(T)}) and (\ref{s(T)}) imply that the energy density of the
radiation bath redshifts as
\begin{equation}
  \frac{\rho_{}^{{\rm rad}}(z_{1})}{\rho_{}^{{\rm rad}}(z_{2})}
  =\frac{g_{\ast}^{}(z_{1})}{g_{\ast}^{}(z_{2})}
  \frac{g_{\ast s}^{4/3}(z_{2})}{g_{\ast s}^{4/3}(z_{1})}
  \left(\frac{1+z_{1}}{1+z_{2}}\right)^{4}.
\end{equation}
In particular, since the standard radiation epoch begins prior to
$z_{c}$, the radiation expanded quasi-adiabatically during the epoch
$z_{c}\geq z\geq z_{eq}$, so we can write:
\begin{equation}
  \label{factor2}
  \frac{\rho_{c}^{{\rm rad}}}{\rho_{eq}^{{\rm rad}}}
  =\frac{g_{\ast}^{}(\,z_{c}^{}\,)}{g_{\ast}^{}(z_{eq}^{})}
  \frac{g_{\ast s}^{4/3}(z_{eq}^{})}{g_{\ast s}^{4/3}(\,z_{c}^{}\,)}
  \left(\frac{1+\,z_{c}^{}\,}{1+z_{eq}^{}}\right)^{4}.
\end{equation}
Note that if we know the phase-space distribution functions describing
each particle species in the radiation bath, then we can compute the
quantities $g_{\ast}^{}(z)$ and $g_{\ast s}^{}(z)$ directly --- again
see Secs. 3.3 and 3.4 in Ref.~\cite{Kolb:1990vq} for more details.
For example, if all relevant particle species are in thermal
  with one another at temperature $T$, then $g_{\ast}=N_{b}
  +(7/8)N_{f}$ and $g_{\ast s}=N_{b}+(7/8)N_{f}$, where $N_{b}$ and
  $N_{f}$ are the total number of relativistic bosonic and fermionic
  degrees of freedom, respectively.

To rewrite the third density ratio appearing in Eq.~(\ref{F_v1c}), note
that conservation of stress-energy ($T^{\mu\nu}_{\;\;\;\;;\nu}=0$) in
the early universe ({\it i.e.} in a spatially-flat FLRW universe)
implies the continuity equation:
\begin{equation}
  \label{continuity}
  \frac{d\rho_{}^{{\rm crit}}}{\rho_{}^{{\rm crit}}}
  =-3[1+\tilde{w}(a)]\frac{da}{a}.
\end{equation}
Here $\tilde{w}(a)$ is the {\it effective} equation-of-state
parameter:
\begin{equation}
  \tilde{w}(a)\equiv w(a)-\frac{8\pi G_{N}\zeta(a)}{H(a)},
\end{equation}
where $w(a)=p(a)/\rho(a)$ is the {\it usual} equation-of-state
parameter [{\it i.e.}\ the ratio of the total cosmological pressure
$p(a)$ to the total cosmological energy density $\rho(a)=\rho_{}^{{\rm
    crit}}(a)$], $H(a)$ is the Hubble expansion rate, and $\zeta(a)$
is the bulk viscosity of the cosmological fluid (see Secs. 2.11 and
15.11 in Ref.~\cite{WeinbergGRbook}).  Integrating Eq.\ 
(\ref{continuity}) from $a_{c}^{}\equiv a(\tau_{c}^{})$ to
$a_{k}^{}\equiv a(\tau_{k}^{})$ yields
\begin{equation}
  \label{continuity_integral}
  \frac{\rho_{k}^{{\rm crit}}}{\rho_{c}^{{\rm crit}}}
  ={\rm exp}\left\{\int_{a_{k}^{}}^{a_{c}^{}}3[1+\tilde{w}(a)]
    \frac{da}{a}\right\}.
\end{equation}
Alternatively, we can define an {\it averaged effective}
equation-of-state parameter $\hat{w}(f)$ through the equation
\begin{equation}
  \label{factor3}
  \frac{\rho_{k}^{{\rm crit}}}{\rho_{c}^{{\rm crit}}}
  =\left(\frac{1+z_{k}^{}}{1+z_{c}^{}}\right)^{3[1+\hat{w}(f)]}.
\end{equation}
Comparing Eqs.\ (\ref{continuity_integral}) and (\ref{factor3}), we
see that $\hat{w}(f)$ is the logarithmic average of the {\it
  effective} equation-of-state parameter $\tilde{w}(a)$ over the range
$a_{k}^{}<a<a_{c}^{}$:
\begin{equation}
  \label{w_hat_from_w_tilde}
  \hat{w}(f)=\frac{1}{{\rm ln}[a_{c}^{}/a_{k}^{}]}
  \int_{a_{k}^{}}^{a_{c}^{}}\tilde{w}(a)\frac{da}{a}.
\end{equation}
Note that if $\tilde{w}(a)$ is an $a$-independent constant over this
range (which we DO NOT assume in this paper) then it becomes equal to
$\hat{w}$.

Finally, we can plug Eqs.\ (\ref{factor1}), (\ref{factor2}),
(\ref{factor3}) into the right-hand-side of Eq.\ (\ref{F_v1c}), solve
for $(1+z_{k}^{})$, and thus find
\begin{equation}
  \label{C1_v2}
  C_{1}(k)=\frac{1}{(1+z_{c}^{})^{2}}\left[\frac{\gamma^{-1/2}}
    {(1+z_{c}^{})}\frac{2\pi f}{H_{0}^{}}\right]^{-4/(1+3\hat{w}(f))}
\end{equation}
where we have defined
\begin{equation}
  \label{def_Gamma_appendix}
  \gamma\equiv\frac{\Omega_{0}^{{\rm mat}}}
  {1\!+\!z_{eq}^{}}\frac{g_{\ast}^{}(\,z_{c}\,)}{g_{\ast}^{}(z_{eq})}
  \frac{g_{\ast s}^{4/3}(z_{eq})}{g_{\ast s}^{4/3}(\,z_{c}\,)}.
\end{equation}

\subsection{Rewriting the factor $\Delta_{h}^{2}(k,\tau_{i})$}
\label{primordial}

Now let us focus on rewriting the factor $\Delta_{h}^{2}(k,\tau_{i})$,
{\it i.e.}\ the primordial tensor power spectrum on short wavelengths.

Recall that the tensor spectral index $n_{t}^{}(k)$ is defined as the
logarithmic slope of the {\it primordial} tensor power spectrum
$\Delta_{h}^{2}(k,\tau_{i})$:
\begin{equation}
  \label{def_n_t}
  n_{t}^{}(k)\equiv\frac{d[{\rm ln}\,\Delta_{h}^{2}(k,\tau_{i})]}
  {d[{\rm ln}\,k]}.
\end{equation}
Integrating this equation, we obtain
\begin{equation}
  \Delta_{h}^{2}(k,\tau_{i})=\Delta_{h}^{2}(k_{{\rm cmb}}^{},\tau_{i})\,
  {\rm exp}\!\left[\int_{k_{{\rm cmb}}^{}}^{k}\!\!n_{t}^{}(k')\frac{dk'}{k'}
  \right],
\end{equation}
where $\Delta_{h}^{2}(k_{{\rm cmb}}^{},\tau_{i})$ is the primordial
tensor power spectrum, evaluated at the CMB ``pivot'' wavenumber
$k_{{\rm cmb}}^{}$.  Alternatively, we can define an {\it averaged}
spectral index $\hat{n}_{t}^{}(f)$ through the equation
\begin{equation}
  \label{def_n_hat_appendix}
  \Delta_{h}^{2}(k,\tau_{i})\equiv\Delta_{h}^{2}(k_{{\rm cmb}}^{},
  \tau_{i})\left[k/k_{{\rm cmb}}^{}\right]^{\hat{n}_{t}^{}(f)}.
\end{equation}
In other words, the {\it effective} spectral index $\hat{n}_{t}^{}(f)$
is nothing but the logarithmic average of the {\it actual} spectral
index $n_{t}^{}(k)$ over the wavenumber range from $k_{{\rm cmb}}^{}$
to $k$:
\begin{equation}
  \label{n_hat_from_n_t}
  \hat{n}_{t}^{}(f)\equiv\frac{1}{{\rm ln}[k/k_{{\rm cmb}}^{}]}
  \int_{k_{{\rm cmb}}^{}}^{k}n_{t}^{}(k')\frac{dk'}{k'}.
\end{equation}
Note that if $n_{t}^{}$ is a $k$-independent constant over this range
(which we DO NOT assume in this paper) then it becomes equal to
$\hat{n}_{t}^{}$.

Finally, it is conventional (and also convenient, for certain
purposes) to define the tensor-to-scalar ratio $r$ through 
the equation:
\begin{equation}
  \label{def_r_appendix}
  r\equiv\frac{\Delta_{h}^{2}(k_{{\rm cmb}}^{},\tau_{i}^{})}
  {\Delta_{{\cal R}}^{2}(k_{{\rm cmb}}^{},\tau_{i}^{})}.
\end{equation}

Combining Eqs.~(\ref{def_n_hat_appendix}) and (\ref{def_r_appendix}),
we can rewrite $\Delta_{h}^{2}(k,\tau_{i})$, the primordial tensor
power spectrum on short wavelengths, in the form:
\begin{equation}
  \label{Delta_h_primordial}
  \Delta_{h}^{2}(k,\tau_{i})=r\,\Delta_{{\cal R}}^{2}(k_{{\rm cmb}}^{},
  \tau_{i})[k/k_{{\rm cmb}}^{}]^{\hat{n}_{t}^{}(f)}.
\end{equation}

\subsection{Recapitulation}
\label{recap}

Now let us assemble our results.  Plugging Eq.~(\ref{C1_v2}) into the
right-hand-side of Eq.~(\ref{T=C1C2C3}), we obtain the tensor transfer
function $T_{h}^{}(k)$ in the useful form:
\begin{equation}
  \label{T_h}
  T_{h}^{}(k)\!=\!\frac{C_{2}^{}(k)C_{3}^{}(k)}{2(1\!+\!z_{c}^{})^{2}}
  \!\left[\frac{\gamma^{-1/2}}{(1\!+\!z_{c})}\frac{2\pi f}{H_{0}^{}}
  \right]^{-4/(1+3\hat{w}(f))},
\end{equation}
Then, plugging Eqs.~(\ref{Delta_h_primordial}) and (\ref{T_h}) into
Eq.~(\ref{master_eq_v1}), we obtain our final result:
\begin{widetext}
  \begin{equation}
    \label{master_eq_v2}
    \Omega_{0}^{{\rm gw}}(f)=\frac{r\Delta_{{\cal R}}^{2}
    (k_{{\rm cmb}}^{},\tau_{i}^{})C_{2}(k)C_{3}(k)\gamma}{24}
    \left(\frac{\gamma^{-1/2}}{(1\!+\!z_{c})}\frac{2\pi f}{H_{0}}
    \right)^{\hat{\alpha}(f)}\left(\frac{a_{0}H_{0}}{k_{{\rm cmb}}^{}}
    \frac{2\pi f}{H_{0}}\right)^{\hat{n}_{t}^{}(f)},
  \end{equation}
\end{widetext}
which is equivalent to Eq.~(\ref{master_eq}) in the text.

In Eqs.~(\ref{T_h}) and (\ref{master_eq_v2}): $C_{2}(k)$ is
given by Eqs.\ (\ref{C2}), (\ref{def_alpha_k}), and
(\ref{def_w_tilde_k}); $C_{3}(k)$ is given by Eqs.~(\ref{C3}) and
(\ref{A}); $\hat{w}(f)$ is given by Eq.~(\ref{w_hat_from_w_tilde});
$\gamma$ is given by Eq.~(\ref{def_Gamma_appendix});
$\hat{n}_{t}^{}(f)$ is given by Eq.~(\ref{n_hat_from_n_t}); and we
have defined
\begin{equation}
  \label{def_alpha_hat_appendix}
  \hat{\alpha}(f)\equiv2\left(\frac{3\hat{w}(f)-1}{3\hat{w}(f)+1}\right).
\end{equation}

Note that, if the quantities $C_{2}(k)$, $C_{3}(k)$, $\hat{w}(f)$ and
$\hat{n}_{t}^{}(f)$ are only weakly $k$-dependent, then the
frequency-dependences of the tensor transfer function $T_{h}^{}(k)$
and the present-day gravitational-wave energy spectrum
$\Omega_{0}^{{\rm gw}}(f)$ are given roughly by
$T_{h}^{}(k)\propto(2\pi f/H_{0})^{-4/(1+3\hat{w}(f))}$ and
$\Omega_{0}^{{\rm gw}}(f)\propto (2\pi f/H_{0})^{\hat{\alpha}(f)
  +\hat{n}_{t}^{}(f)}$, respectively.

\section{Deriving the frequencies $f_c$ and $f_{\rm end}$}
\label{f_c_f_end_derivation}

In this appendix we derive Eqs.~(\ref{f_c}) and (\ref{f_end}), for
$f_{c}^{}$ and $f_{{\rm end}}^{}$, respectively.  Let us start with
Eq.~(\ref{f_c}) for $f_{c}$.  We start by writing
\begin{subequations}
  \begin{eqnarray}
    \label{f_c_v1}
    (2\pi f_{c})^{2}&=&\frac{a_{c}^{2}}{a_{0}^{2}}
    \frac{H_{c}^{2}}{H_{0}^{2}}H_{0}^{2}\\
    \label{f_c_v2}
    &=&\frac{H_{0}^{2}}{(1+z_{c})^{2}}
    \frac{\rho_{c}^{{\rm crit}}}{\rho_{0}^{{\rm crit}}} \\
    \label{f_c_v3}
    &=&\frac{H_{0}^{2}}{(1+z_{c})^{2}}
    \frac{\rho_{eq}^{{\rm mat}}}{\rho_{0}^{{\rm crit}}}
    \frac{\rho_{c}^{{\rm rad}}}{\rho_{eq}^{{\rm rad}}}.
  \end{eqnarray}
\end{subequations}
If these steps are unclear, see Eqs.~(\ref{F_v1a}), (\ref{F_v1b}),
(\ref{F_v1c}) and the paragraph that follows them.  Now, substituting
Eqs.~(\ref{factor1}) and (\ref{factor2}) into Eq.~(\ref{f_c_v3}), and
solving for $f_{c}$, we obtain Eq.~(\ref{f_c}) as desired.

Next let us derive Eq.~(\ref{f_end}) for $f_{{\rm end}}^{}$.  To
begin, note that we can write $\Delta_{h}^{2}(k_{{\rm
    end}}^{},\tau_{i})$ in two different ways.  On the one hand, using
Eqs.~(\ref{def_n_hat_appendix}) and (\ref{def_r_appendix}), we can
write
\begin{subequations}
  \begin{equation}
    \label{Delta_h_v1}
    \Delta_{h}^{2}(k_{{\rm end}}^{},\!\tau_{i}^{})=
    r\Delta_{{\cal R}}^{2}(k_{{\rm cmb}}^{},\!\tau_{i}^{})\!
    \left[\frac{2\pi f_{{\rm end}}^{}}{k_{{\rm cmb}}^{}/a_{0}^{}}
    \right]^{\hat{n}_{t}^{}(f_{{\rm end}}^{})}\!\!.
  \end{equation}
  On the other hand, we can use the well-known inflationary formula
  \begin{equation}
    \label{Delta_h_v2}
    \Delta_{h}^{2}(k_{{\rm end}}^{},\tau_{i})=
    64\pi G_{N}^{}\left(\frac{H_{{\rm end}}^{}}{2\pi}\right)^{2},
  \end{equation}
\end{subequations}
where our conventions match those of the WMAP experiment ({\it e.g.}\ 
see Eq.~(A13) in Ref.~\cite{Peiris:2003ff}).  Comparing
Eqs.~(\ref{Delta_h_v1}) and (\ref{Delta_h_v2}), we see that
\begin{equation}
  \label{H_end}
  H_{{\rm end}}^{2}=\frac{\pi^{2}r\Delta_{{\cal R}}^{2}
    (k_{{\rm cmb}}^{},\tau_{i})}{16\pi G_{N}^{}}
  \left(\frac{2\pi f_{{\rm end}}^{}}{k_{{\rm cmb}}^{}/a_{0}^{}}
  \right)^{\hat{n}_{t}^{}(f_{{\rm end}}^{})}.
\end{equation}
Next, from the definition of $f_{{\rm end}}^{}$, we can write
\begin{equation}
  \label{f_end_v1}
  (2\pi f_{{\rm end}}^{})^{2}=\frac{a_{{\rm end}}^{2}}{a_{0}^{2}}
  H_{{\rm end}}^{2}.
\end{equation}
Note that, since the first factor on the right-hand-side of
Eq.~(\ref{f_end_v1}) is nothing but $C_{1}(k_{{\rm end}}^{})$, we
can use Eq.~(\ref{C1_v2}) to rewrite it as:
\begin{equation}
  \label{C1_end}
  \frac{a_{{\rm end}}^{2}}{a_{0}^{2}}=\frac{1}{(1+z_{c}^{})^{2}}
  \left[\frac{\gamma^{-1/2}}{(1+z_{c}^{})}\frac{2\pi f_{{\rm end}}^{}}
    {H_{0}^{}}\right]^{-4/(1+3\hat{w}(f_{{\rm end}}^{}))}.
\end{equation}
Finally, we can plug Eqs.~(\ref{H_end}) and (\ref{C1_end}) into the
right-hand-side of Eq.~(\ref{f_end_v1}), and solve for $f_{{\rm
    end}}^{}$ to obtain Eq.~(\ref{f_end}) as desired.

\section{Numerical formulae}
\label{numbers}

This appendix lists a few results that are useful for converting our
algebraic expressions into numerical results and plots.

At matter-radiation equality, we have the standard values
$g_{\ast}^{}(z_{eq})=3.3626$ and $g_{\ast s}^{}(z_{eq})=3.9091$; and
at BBN (when the temperature is $T=1~{\rm MeV}$) we have the standard
values $g_{\ast}^{}(z_{{\rm bbn}}^{})=g_{\ast s}^{}(z_{{\rm
    bbn}}^{})=10.75$ \cite{Kolb:1990vq}.  From WMAP 3rd year data
alone, we know $\Delta_{{\cal R}}^{}(k_{{\rm
    cmb}}^{})=(2.04\pm0.14)\times10^{-9}$; and from WMAP 3rd year data
plus lyman-alpha-forest data we know $\Delta_{{\cal R}}^{}(k_{{\rm
    cmb}}^{})=(2.24\pm0.11)\times10^{-9}$ \cite{Pat}, where these
values for $\Delta_{{\cal R}}^{}(k_{{\rm cmb}}^{})$ are quoted at
$k_{{\rm cmb}}^{}=0.05~{\rm Mpc}^{-1}$.  The present day value of the
Hubble expansion rate may be written as $H_{0}^{}=(3.24)h\times
10^{-18}{\rm Hz}$, where the Hubble parameter $h\approx0.72$ is a
fudge factor that absorbs the uncertainty in the measurement of
$H_{0}^{}$.  Thus we can write $(2\pi f/H_{0})=(1.94/h)\times10^{18}
(f/{\rm Hz})$.  If $k_{{\rm cmb}}^{}=0.002~{\rm Mpc}^{-1}$, then
$(k_{{\rm cmb}}^{}/a_{0}H_{0})=6.00/h$; and if $k_{{\rm cmb}}^{}=
0.05~{\rm Mpc}^{-1}$, then $(k_{{\rm cmb}}^{}/a_{0}H_{0})=150.0/h$.
The redshift of matter-radiation equality may be written as
$(1+z_{eq}^{})=2.3\times10^{4}\Omega_{0}^{{\rm mat}}h^{2}$; and
the redshift of BBN (when the temperature was $T=1~{\rm MeV}$) 
is $(1+z_{{\rm bbn}}^{})=5.9\times10^{9}$.

\end{document}